\documentclass[sigconf]{acmart}

\usepackage{bm}         

\usepackage{graphicx}   
\usepackage{float}      
\usepackage{wrapfig}    
\usepackage{subcaption} 

\usepackage{multirow}   
\usepackage{makecell}   
\usepackage{colortbl}   

\usepackage{algorithm}

\usepackage{xcolor}
\definecolor{grice}{HTML}{ddd5c3} 
\definecolor{gblue}{HTML}{78a5ab} 

\usepackage{url}

\AtBeginDocument{%
  \providecommand\BibTeX{{%
    Bib\TeX}}}

\setcopyright{acmlicensed}
\copyrightyear{2018}
\acmYear{2018}
\acmDOI{XXXXXXX.XXXXXXX}
\acmConference[Conference acronym 'XX]{Make sure to enter the correct
  conference title from your rights confirmation email}{June 03--05,
  2018}{Woodstock, NY}
\acmISBN{978-1-4503-XXXX-X/2018/06}

\copyrightyear{2025}
\acmYear{2025}
\setcopyright{acmlicensed}\acmConference[RecSys '25]{Proceedings of the Nineteenth ACM Conference on Recommender Systems}{September 22--26, 2025}{Prague, Czech Republic}
\acmBooktitle{Proceedings of the Nineteenth ACM Conference on Recommender Systems (RecSys '25), September 22--26, 2025, Prague, Czech Republic}
\acmDOI{10.1145/3705328.3748069}
\acmISBN{979-8-4007-1364-4/2025/09}
\begin{document}

\title{Paragon: Parameter Generation for Controllable Multi-Task Recommendation}


\author{Chenglei Shen}
\authornote{Both authors contributed equally to this research.}
\affiliation{%
  \institution{Gaoling School of Artificial Intelligence, Renmin University of China}
  \city{Beijing}
  \country{China}
  }
\email{chengleishen9@ruc.edu.cn}

\author{Jiahao Zhao}
\authornotemark[1]
\affiliation{%
  \institution{Gaoling School of Artificial Intelligence, Renmin University of China}
  \city{Beijing}
  \country{China}
  }
\email{zhaojiahao2202@ruc.edu.cn}

\author{Xiao Zhang}
\authornote{Corresponding author: Xiao Zhang (e-mail: zhangx89@ruc.edu.cn).}
\affiliation{%
  \institution{Gaoling School of Artificial Intelligence, Renmin University of China}
  \city{Beijing}
  \country{China}
  }
\email{zhangx89@ruc.edu.cn}

\author{Weijie Yu}
\affiliation{%
  \institution{School of Information Technology
and Management, University of International Business and Economics}
  \city{Beijing}
  \country{China}
  }
\email{yu@uibe.edu.cn}

\author{Ming He}
\affiliation{%
  \institution{AI Lab at Lenovo Research}
  \city{Beijing}
  \country{China}
}
\email{heming01@foxmail.com}

\author{Jianping Fan}
\affiliation{%
  \institution{AI Lab at Lenovo Research}
  \city{Beijing}
  \country{China}
}
\email{jfan1@lenovo.com}

\renewcommand{\shortauthors}{Trovato et al.}
\renewcommand{\shorttitle}{Paragon: Parameter Generation for Controllable Multi-Task Recommendation}

\begin{abstract}
Commercial recommender systems face the challenge that task requirements from platforms or users often change dynamically (e.g., varying preferences for accuracy or diversity). Ideally, the model should be re-trained after resetting a new objective function, adapting to these changes in task requirements. However, in practice, the high computational costs associated with retraining make this process impractical for models already deployed to online environments. 
This raises a new challenging problem: how to efficiently adapt the learned model to different task requirements by controlling the model parameters after deployment, without the need for retraining.
To address this issue, we propose a novel controllable learning approach via \textbf{para}meter \textbf{g}eneration for c\textbf{on}trollable multi-task recommendation (\textbf{Paragon}), which allows the customization and adaptation of recommendation model parameters to new task requirements without retraining. 
Specifically, we first obtain the optimized model parameters through adapter tunning based on the feasible task requirements. Then, we utilize the generative model as a parameter generator, employing classifier-free guidance in conditional training to learn the distribution of optimized model parameters under various task requirements. Finally, the parameter generator is applied to effectively generate model parameters in a test-time adaptation manner given task requirements. 
Moreover, Paragon seamlessly integrates with various existing recommendation models to enhance their controllability. Extensive experiments on two public datasets and one commercial dataset demonstrate that Paragon can efficiently generate model parameters instead of retraining, reducing computational time by at least 94.6\%.
The code is released at \href{https://github.com/bubble65/Paragon}{https://github.com/bubble65/Paragon}.
\end{abstract}


\ccsdesc[300]{Information systems~Recommender systems}






\maketitle

\begin{figure}[h]
    \centering
    \begin{subfigure}[b]{0.42\textwidth}
        \centering
        \includegraphics[width=\textwidth]{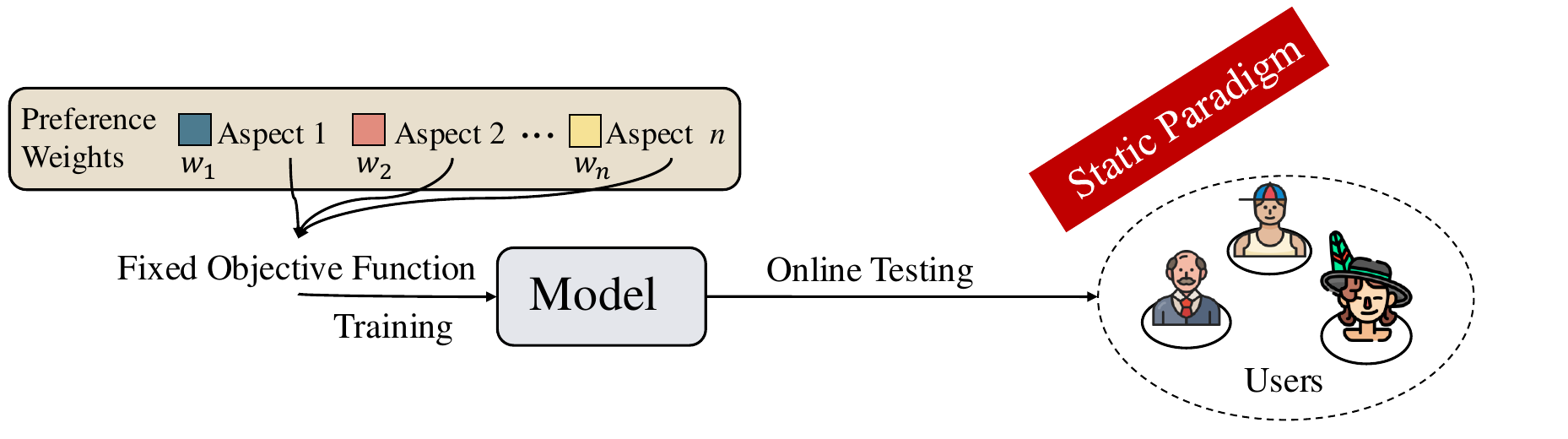}
        \caption{Multi-task recommendation model in static environment. }
    \end{subfigure}
    \begin{subfigure}[b]{0.42\textwidth}
        \centering
        \includegraphics[width=\textwidth]{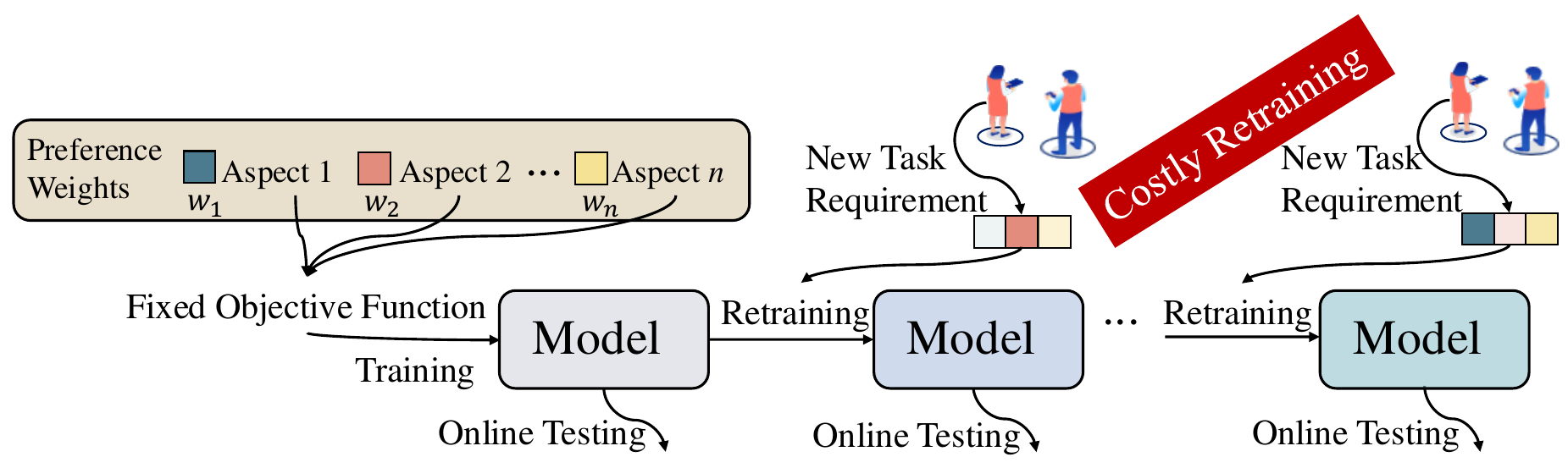}
        \caption{Multi-task recommendation model in dynamic environment.}
    \end{subfigure}
    \caption{Illustration of multi-task recommendation in static and dynamic environments. ``Aspects'' represents the aspects of recommended results such as diversity, fairness, etc. ``Preference weights'' is a vector of preferences across different aspects. Each new task requirement can be represented by a vector of novel preference weights (maybe different).}
    \label{fig:traditional_paradigms}
    \vspace{-5mm} 
\end{figure}

\section{Introduction}
\label{sec:intro}



Traditional recommender systems focus on improving accuracy through analyzing user behaviors and contextual data~\citep{kang2018self,hidasi2016parallel,zhang2024model,zhang2024saqrec,zhang2025test, zhang2024reinforcing, zhang2025comment, zhang2024modeling}. 
Nowadays, recommendation, especially multi-task recommendation (MTR), places greater emphasis on multiple aspects like diversity and fairness~\citep{xia2017adapting,oosterhuis2021computationally}. They simultaneously optimize multiple aspects based on the task requirements (i.e., preference weights for each aspect). This has made MTR a research hotspot by leveraging unified models to learn interrelated aspects for mutual improvement~\cite{he2022metabalance,yang2023adatask,su2024stem,wang2023multi}.

Nonetheless, in the online stage of recommendation, there is often a need for sudden, instantaneous changes in task requirements, where the preference weights for each aspect can change unexpectedly. From a commercial perspective, businesses often need real-time adjustments to their recommendation strategies, especially during live events like Black Friday, where user interests can shift drastically in a matter of hours. From a user’s perspective, preferences can change unexpectedly as well. For instance, a user might initially prioritize highly accurate recommendations when searching for a product, but after several similar items, they may start preferring more diverse suggestions to explore new options. These changes are sudden and instantaneous during the online testing stage, making it impossible to predict them in a data-driven manner, highlighting the importance of considering sudden changes in task requirements in MTR to avoid sub-optimal recommendations. 

\begin{figure}[t]
    \centering
    \includegraphics[width=0.93\linewidth]{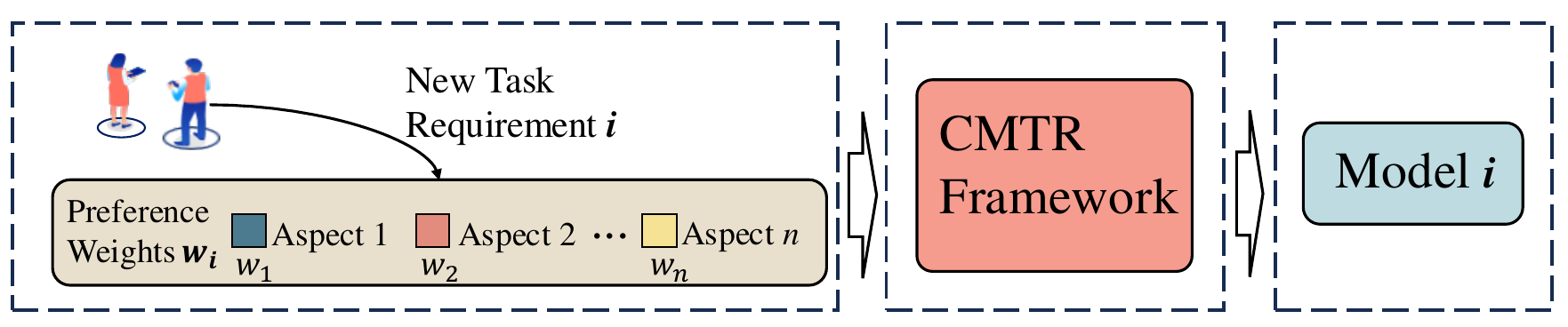}
     \caption{Illustration of controllable multi-
task recommendation (CMTR) given task requirement $i$ at testing time.}
    \label{fig:new_paradim}
    \vspace{-5mm}
\end{figure}

Traditional MTR algorithms face challenges in addressing this issue (i.e., the sudden changing preference weights on each aspect). Most existing MTR algorithms focus on achieving mutual improvement in \emph{static} environment. As shown in Figure~\ref{fig:traditional_paradigms} (a), the task requirement (the preference weights on each aspect such as accuracy and diversity) are predefined and fixed during both training and testing~\cite{zhang2021survey,sener2018multi}. When traditional MTR algorithms are deployed in the \emph{dynamic} environment, they require reconstructing the objective function, retraining the model, and redeploying it to the online system whenever a new task requirement arises, as depicted in Figure~\ref{fig:traditional_paradigms} (b). Obviously, when faced with a sudden change in task requirements, it cannot respond immediately due to the highly time- and resource-intensive nature of retraining. Overall, the central issue is that task requirements are sudden changing and manually specified at test time, resulting in unavoidable retraining overhead for new demands. Consequently, developing efficient methods to adjust models to real-time task requirements is crucial, which we term ‌controllable multi-task recommendation (CMTR)‌.





CMTR aims to develop an efficient test-time model control mechanism that eliminates the need for retraining, as illustrated in Figure~\ref{fig:new_paradim}. In fact, retraining essentially involves adjustments to the model parameters. To capture these global parameter changes, we innovatively use a generative model to generate recommendation model parameters instead of relying on the time-consuming retraining process. Additionally, to preserve the foundational capabilities of the recommendation model and reduce the learning complexity and inference time of the generative model, we generate only task-specific adapters. Moreover, treating the dynamically changing task requirements as conditioning signals  further confers test‑time controllability on CMTR, thereby eliminating the need to retrain the recommendation model during inference. 
Specifically, we proposed Paragon, begins by constructing an objective function aligned with task-specific preference weights, and through advanced optimization techniques, we fine-tune recommendation model parameters using adapter tuning. We then train a generative model (e.g., diffusion model) to learn the conditional distribution of these optimized adapter parameters under various task requirements, where the classifier-free guidance training strategy is employed to perform conditional training. Once trained, during online testing, the generative model can rapidly generate task-specific adapter parameters conditioned on the task requirements, which are then integrated with the backbone to produce recommendation lists that meet the specified requirements. Extensive experiments demonstrate that: (1) Paragon can rapidly and instantly generate high-performance model parameters without retraining. (2) The parameters generated by Paragon exhibit strong robustness. (3) Paragon performs well in controlling multiple aspects beyond accuracy and diversity.

We summarize our contributions as follows: 

\begin{itemize}
    \item   
  We reveal the impact of sudden changes in task requirements during online testing on MTR and define the CMTR task, which emphasizes the model's adaptability to sudden changing task requirements in online scenarios.
    \item 
    We introduce Paragon, which generates recommendation model parameters based on task requirements, eliminating costly retraining.   
    \item Extensive experiments show that Paragon cuts computational time by at least 94.6\% while preserving virtually the same recommendation performance as retraining.
\end{itemize}

\section{Related Work}
\label{gen_inst}






\textbf{Multi-task learning (MTL)} aims to develop unified models that tackle multiple learning tasks simultaneously while facilitating information sharing \citep{zhang2021survey,ruder2017overview,shi2024unisar,shi2025unified,shen2024survey}. Recent advancements in MTL include deep networks with various parameter sharing mechanisms~\citep{misra2016cross,long2017learning,yang2016deep} and approaches treating MTL as a multi-objective optimization problem~\citep{lin2019pareto, mahapatra2020multi, xie2021personalized}. These latter methods focus on identifying Pareto-efficient solutions across tasks, with significant applications in recommender systems \citep{jannach2022multi, li2020video, zheng2022survey}. Researchers have explored different strategies, from alternating optimization of joint loss and individual task weights to framing the process as a reinforcement learning problem \citep{xie2021personalized}. The emphasis has shifted from optimizing specific preference weights to finding weights that achieve Pareto efficiency across objectives \citep{sener2018multi, lin2019pareto,Liu2019attention}. Recently approaches, such as the CMR~\citep{Chen2023cmr}, utilize hypernetworks to learn the trade-off curve for MTL problems. However, our novel approach diverges from these existing methods by employing diffusion models to control model parameters at test time, potentially offering greater flexibility and adaptability in handling multi-task learning problems. Recently, some studies have utilized large language models (LLMs) to influence recommendation systems~\citep{maps, vaps} or act as recommenders~\citep{more, lrgd}. However, these methods are unable to provide flexible parameter control and instead depend on semantic understanding, a characteristic that fundamentally differentiates them from our approach. Moreover, there has been a growing body of research focusing on dynamic model adaptation at test time~\cite{shen2023hyperbandit,Chen2023cmr, shen2024survey}.

\textbf{Diffusion models.}
Diffusion probabilistic models~\citep{ho2020denoising, song2020denoising, nichol2021improved}have not only achieved significant success in the field of image generation but have also found wide applications in various other areas in recent years, such as video generation ~\citep{Ho2022video}, text generation~\citep{li2022diffusion,gong2022diffuseq}, etc.  Moreover, diffusion models have shown the ability to generate high-quality neural network parameters, achieving comparable or even superior performance to traditionally trained models~\citep{yuan2024spatio,schurholt2022hyper,knyazev2021parameter,wang2024neural}. These models have also been applied to enhance the accuracy of recommender systems by addressing challenges such as noisy interactions and temporal shifts in user preferences~\citep{wang2023diffusionrec}. In our work, we utilize diffusion models to generate parameters for controllable multi-task recommender systems.  Notably, we use generative models to generate high-quality parameters for recommendation models. The parameter generation paradigm differs from generative retrieval~\citep{gear,luc}, which directly generates item identifiers. Our approach dynamically reconfigures models, maintaining efficiency and enabling control.



\section{Problem Formulation and Analyses} 
\label{sec:formulation}
Given a user $u\in \mathcal{U}$ and a set of candidate items  $\mathcal{C} = \{c_k\}_{k=1}^{|\mathcal{C}|}$ where $|\mathcal{C}|$ denotes the total number of candidate items. The historical interaction sequence of user $u$ of length $h$ is denoted by $S_{u} = \{c_1^{u}, 
c_2^{u}, \dots, c_{h}^{u}\}$ (also called user history), where $c_k^{u} \in \mathcal{C}, k \in \{1, 2, \ldots, h\}$. 
For a \textbf{recommendation task} $i \in \{1, 2, \ldots, N\}$, a recommender system aims to find the following item list $L_{i}^*$ among all possible lists $\{L\}$ composed by candidate items from $\mathcal{C}$: 
\begin{equation}
\label{eq:MTL:total_reward}
L_i^*=\underset{L}{\arg\max}~ R_i (L ~|~ S^u, \mathcal{C} ),
\end{equation}
where $R_i$ denotes the reward function corresponding to task $i$, which evaluates the recommender system's performance with respect to task $i$. 
More specifically, modern recommender systems often evaluate performance from multiple perspectives, the reward function in Eq.~(\ref{eq:MTL:total_reward}) for task $i$ can be expressed as the following linear combination of $p$ utility functions $\{ U_j\}_{j=1}^p$: 
\begin{equation}
\label{eq:MTL:weight_utility}
R_i (L(S_{u}, \mathcal{C}))=\sum_{j=1}^{p} w_i^j~ U_j (L ~|~ S_{u}, \mathcal{C}),
\end{equation}
which allows task $i$ to be quantified  by a set of \textbf{preference weights} $\bm w_i = \{ w_i^j \}_{j=1}^{p} \in \mathcal{W}$ for the various utilities, where $\mathcal{W}$ denotes the preference weight space that is a simplex. 

Then, we can provide the definition of \textbf{controllable multi-task recommendation (CMTR)}. 
The goal of CMTR is to find a recommendation model $f_{\bm \theta}$, parameterized by $\bm \theta \in \Theta$, such that the item lists output during test time, $L = f_{\bm \theta}(S_{u}, \mathcal{C})$, can adapt to changes in tasks (i.e., adapt to variations in the corresponding preference weights in Eq.~(\ref{eq:MTL:weight_utility})).
As an example, after the recommendation model $f_{\bm \theta}$ is deployed, when the preference weights for different utilities (e.g., accuracy and diversity) need to shift from $\bm w_i = \{ w_i^j \}_{j=1}^{p}$ (i.e., task $i$) to $\bm w_k =  \{ w_k^j \}_{j=1}^{p}$ (i.e., task $k$) based on user or platform requirements, we say that the recommendation model $f_{\bm \theta}$ is \textbf{controllable} if it can ensure that its reward remains at a high level regardless of how the preference weights change.
Ideally, to accommodate changes in tasks, we could retrain the recommendation model after receiving new preference weights to update its parameters, resulting in $f_{\tilde{\bm \theta}}$ that maintains a high reward. However, for an already deployed model, the time required for retraining is impractical and unacceptable. 
Another straightforward method would be to store $N$ sets of task-specific parameters corresponding to the preference weights for $N$ tasks at the time of deployment, and load them when a new task arises at test time. 
However, when considering a continuous preference weight space where the number of tasks $N$ tends to infinity (i.e., a continuous task space), this discrete method becomes impractical due to storage limitations and cannot accommodate fine-grained or continuous task variations.

To efficiently and effectively adapt to changes in tasks, this paper focuses on controlling the model parameters $\bm \theta$ of the recommendation model \(f_{\bm \theta}\) to accommodate the varying preference weights of new tasks. Specifically, we treat the preference weights as variables and model the relationship between the preference weight space \(\mathcal{W}\) and the model parameter space \(\Theta\) during training, transforming the time- and resource-intensive retraining problem at test time into an efficient inference problem. 
Formally, we aim to find a function \(g_{\bm \xi}: \mathcal{W} \rightarrow \Theta\) (where $\bm \xi $ denotes the parameter of $g$) that generates model parameters capable of achieving a high reward given the new preference weights \(\bm w_k\) for any task \(k\) at test time:
\begin{equation}
\label{eq:MTL:weight_utility:diffusion}
R_k (L(S_{u}, \mathcal{C}))=\sum_{j=1}^{p} w_k^j~ U_j (f_{\bm \theta_k}( S_{u}, \mathcal{C}) ~|~ \bm \theta_k = g_{\bm \xi}(\bm w_k)).
\end{equation} 
In contrast to traditional multi-task recommendation (MTR), which focuses only on \emph{fixed} preference weights for different utilities, our defined CMTR emphasizes how the model adapts to \emph{dynamic} changes in preference weights after deployment.  
This shift means that in traditional MTR, each task corresponds to a \emph{single} utility, whereas in CMTR, each task is associated with \emph{multiple} utilities combined through a linear weighting, with combination coefficients determined by a set of task-specific preference weights. 
As a result, CMTR places greater emphasis on test-time adaption to handle dynamic task requirements, introducing new challenges for CMTR model training and construction compared to MTR.

\begin{figure*}[t]
    \centering
    \includegraphics[width=0.98\textwidth]{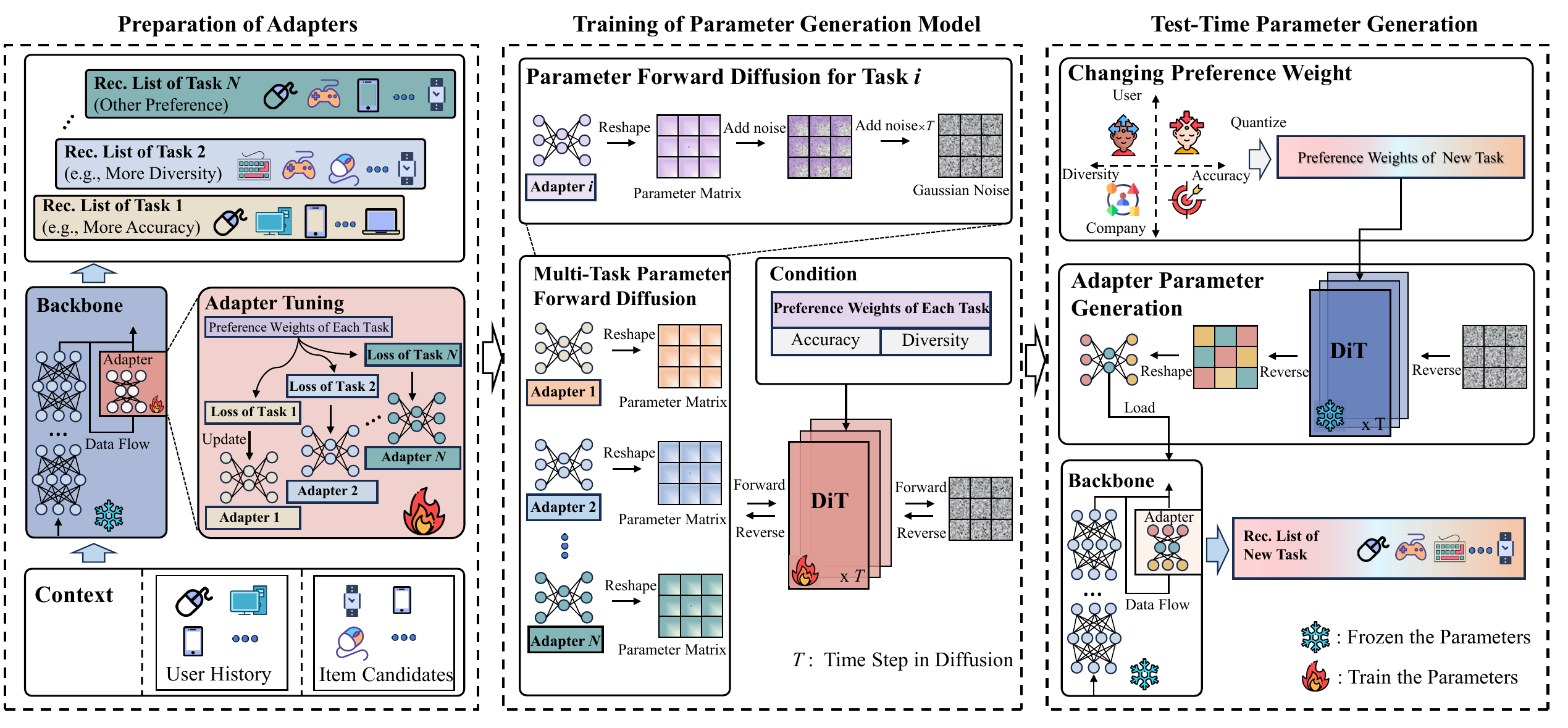}
    \caption{An overview of the proposed Paragon.  }
    \label{fig:overview}
    \Description{An overview of the proposed Paragon.}
\end{figure*}

\section{Paragon: The Proposed Approach}
\label{method}

In this section, we provide a detailed description of the proposed approach, \textbf{Paragon}. 
As shown in Figure~\ref{fig:overview}, we provide an illustrative overview of the proposed Paragon, which contains the following three phases. 
(1)~\emph{Preparation of adapters}: 
the left part in Figure~\ref{fig:overview} shows the training process of the recommendation model, from which we can obtain a collection of optimized adapter parameters for feasible specific task by sampling the preference weights. We focus on two utilities: accuracy and diversity. As defined in Eq.~(\ref{eq:MTL:weight_utility:diffusion}), each task is represented by a set of preference weights for these two utilities.(2)~\emph{Parameter diffusion model training}: 
the middle part in Figure~\ref{fig:overview} illustrates the conditional training procedure of the generative model $g_{\bm\xi}$ (i.e., DiT) with the optimized adapter parameters as initial data and the corresponding preference weights as condition, thus generating meaningful adapter parameters from Gaussian noise given preference weights.  
(3)~\emph{Test-time parameter generation}: 
the right part in Figure~\ref{fig:overview} shows how we utilize the trained DiT model at test time to adapt to dynamically changing task requirements (i.e., preference weights for diversity and accuracy). First, we quantify these task requirements as preference weights.  Next, we employ the trained DiT model to generate adapter parameters at test time, using these preference weights as inputs, which are then combined with the backbone to directly support the recommendation task.

\subsection{Preparation of Adapters}
\label{sec:preparation}
Our goal is to construct the parameters of optimized recommendation models under different preference weights to prepare data for the generative model. Thus, this section is organized into three parts: the structure of the recommendation model, the construction of task-specific objective functions, and the tuning process for the recommendation model parameters.

\noindent \textbf{Model structure.}
As shown in the left module of Figure~\ref{fig:overview}, sequential recommendation models take user history and candidate items as input. Guided by the objective function (i.e., loss function), the model learns the underlying relationships within the user history, ultimately generating a recommendation list (i.e., Rec. List) from the candidate items. To preserve the foundational capabilities of the recommendation model and reduce the learning complexity and inference time of the generative model. we adopt a backbone recommendation model and task-specific adapters framework, where the backbone is trained to preserve the accurate recommendation by BPR loss while the adapters are trained under various task requirements for generative model to learn. Specifically, we incorporate the adapter using a residual connection, attaching it to the last layer of the backbone model. Based on this structure, the backbone is fixed after training, and on this foundation, the task-specific adapters are independently trained under the guidance of their respective objective functions. 

\noindent\textbf{Objective function construction.}
To obtain the optimized
 task-specific adapter parameters under CMTR setting (as shown in Sec.~\ref{sec:formulation} ), we first focus on the construction of loss functions based on different preference weights of each task. Specifically, we directly convert the reward maximization problem (reward defined in Eq.~(\ref{eq:MTL:weight_utility})) into a loss minimization problem. Given a specific set of preference weights $\bm w_i = \{ w_i^j \}_{j=1}^{p} \in  \mathcal{W}$, which represent preference weight for the $j$-th utility in the requirement of task $i$.  
 Here, we focus on two utilities including diversity loss $\ell_{\text{diversity}}$ and accuracy loss $\ell_{\text{accuracy}}$ in each task (i.e., $p=2$). Thus, the total loss function for task $i$ is
 \begin{equation}
 \label{eq:diffusion:total_loss:p=2}
\ell_{i} = 
w_i^1\ell_{\text{accuracy}}
+ w_i^2\ell_{\text{diversity}},
\end{equation}
where $\ell_{\text{accuracy}}$ employs the BPR loss~\cite{rendle2012bpr}, and the formulation of $\ell_{\text{diversity}}$ is detailed in the next section.

\noindent \textbf{Adapter tuning.} Based on above total loss function, we decompose the recommendation model parameters $\bm\theta$ into two components: task-specific adapter parameters, denoted as  $\bm \theta_{\text{a}}$
and task-independent backbone parameters, denoted as $\bm \theta_{\text{b}}$. 
  Accordingly, optimizing the model is divided into two phases. The \emph{first phase} focuses on optimizing the backbone parameters $\bm \theta_{\text{b}}$, which uses the standard BPR loss to train the backbone model thus preserving the original recommendation accuracy. The \emph{second phase} is about the optimization of the task-specific adapter parameters $\bm \theta_{\text{a}}$, which aims at improving the system’s adaptability to different tasks. 
  During the second phase, the backbone parameters are frozen to prevent them from being tailored to any specific task, whereas the adapter is trainable.  
  More specifically, in the second phase, we train the task-specific adapter parameters based on two loss functions as in Eq.~(\ref{eq:diffusion:total_loss:p=2}), one for accuracy and one for diversity.
For the accuracy loss $\ell_\text{accuracy}$, we continue to use BPR as the loss function to guide the model toward accuracy. For the diversity loss $\ell_{\text{diversity}}$, inspired by \citet{yan2021diversification}, we apply a differentiable smoothing of the $\alpha$-DCG metric and adapt it to the recommendation setting. Consider $|\mathcal{C}|$ candidate items and $|\mathcal{M}|$ categories, where each item may cover 0 to $|\mathcal{M}|$ categories. The category labels are denoted as  $y_{k,l}$: $y_{k,l} = 1$ if item $k$ covers category $m$, and $y_{k,l} = 0$ otherwise, where $k\in \{0,\ldots, |\mathcal{C}|-1\}$, $l\in \{0,\ldots, |\mathcal{M}|-1\}$. Based on the $\alpha$-DCG, we design a differentiable diversity loss function:
\begin{equation}
\ell_{\text{diversity}} = 
-  \sum_{k=1}^{|\mathcal{C}|} \sum_{l=1}^{|\mathcal{M}|} \frac{y_{k,l} (1 - \alpha) C_{k,l}}{\log_2(1 + \text{Rank}_k)},
\end{equation}
where $\alpha$ is a hyper parameter between 0 and 1, $\text{Rank}_k$ is the soft rank of the item $k$, and $C_{k,l}$ is the number of times the category $l$ being covered by items prior to the soft rank $\text{Rank}_k$. That is:
\begin{equation}
\begin{aligned}
\text{Rank}_k &= 1 + \sum_{j \neq k} \text{sigmoid} \left( (s_j - s_k) / T \right),\\
\quad 
C_{k,l} &= \sum_{j \neq k} y_{j,l} \cdot \text{sigmoid} \left( (s_j - s_k) / T \right),
\end{aligned}
\end{equation}
where $s_k$ denotes the relevance score of the $k$-th candidate item output by the model. For  task $i$, we denote $\bm \theta_i$ as the model parameters including task-specific adapter parameters $\bm \theta_i^{\text{a}}$ and fixed backbone parameters $\bm \theta_i^{\text{b}}$. Based on the total loss in Eq.~(\ref{eq:diffusion:total_loss:p=2}), the task-specific optimization process of $\bm \theta_i^{\text{a}}$ for task $i$ can be formulated as follows:
\begin{equation}
\label{eq:diffusion:total:spec}
\bm \theta_i^{\text{a}} = \operatornamewithlimits{arg\,min}_{\bm \theta^{\text{a}}_i} ~
w_i^1\ell_{\text{accuracy}}
+ w_i^2\ell_{\text{diversity}},
\end{equation}
where $\bm w_i = \{ w_i^1, w_i^2\} \in  \mathcal{W}$ is sampled from $[0,1]$. We employ the standard Adam optimizer to optimize these parameters. Then we transform the parameters of each task-specific adapter into a matrix-based format and these optimized parameters serve as the ground truth for the subsequent generative model training process. 

\subsection{Training of Parameter Generation Model}
The optimized adapter parameters and corresponding preference weights obtained from Sec.~\ref{sec:preparation} are used as the training data for the diffusion model. We employ a generative model $g_{\bm \xi} $ parameterized by $\bm \xi$ to learn the process of generating model parameters. 
Specifically, $g_{\bm \xi} $ is applied to predict the conditional distribution of the adapter parameter matrices $p_{g_{\bm \xi}}(\bm \theta_i^\text{a}|\bm w_i)$ given the preference weights $\bm w_i$, where $i$ corresponds to the task $i$.  We adopt diffusion models~\citep{ho2020denoising} as our generative model due to its efficacy in various generation tasks~\citep{li2022diffusion, ho2022imagen,vignac2023digress} and its superior performance on multi-modal conditional generation~\citep{bao2023one,nichol2022glide,saharia2022photorealistic}. We train the diffusion model to sample parameters by gradually denoising the optimized adapter parameter matrix from the Gaussian noise.  This process is intuitively reasonable as it intriguingly mirrors the optimization journey from random initialization which is a well-established practice in existing optimizers like Adam. For task $i$, our denoising model takes two parts as the input: a noise-corrupted adapter parameter matrix $\bm \theta_{i,t}^\text{a}$, and a set of preference weights $\bm w_i$, with $t$ representing the step in the forward diffusion process.  The training objective is  as follows:
\begin{equation}
    \ell_{\text{diff}} = \mathbb{E}_{\bm \theta_{i,0}^\text{a},\epsilon\sim\mathcal{N}(0,1),t}\left[\left\|\epsilon-\epsilon_{\bm \xi}(\bm{\theta}_{i,t}^\text{a},\bm w_i, t) \right\|^2 \right], 
\end{equation}
where $\epsilon$ denotes the noise to obtain $\bm \theta_{i,t}^\text{a}$ from $\bm \theta_{i,0}^\text{a}$, and the denoising model $\epsilon(\cdot)$ is the main part of the generative model $g_{\bm \xi}$. We assume that the parameters of $g_{\bm \xi}$ primarily originate from the denoising model. For simplicity, we denote the denoising model as $\epsilon_{\bm \xi}$.  To conduct condition training in a classifier-free guidance manner~\citep{ho2022classifier}, we use the denoising model to serve as both the conditional and unconditional model by simply inputting a null token $\varnothing$ as the condition (i.e., preference weights $\bm w_i$) for the unconditional model, i.e.\ $\epsilon_{\bm \xi}(\bm\theta_{i,t}^\text{a},t) = \epsilon_{\bm \xi}(\bm \theta_{i,t}^\text{a}, \bm w_i = \varnothing, t)$. The probability of setting $\bm{w}_i$ to $\varnothing$ is denoted as $p_\mathrm{uncond}$ and is configured as a hyperparameter.

\subsection{Test-Time Parameter Generation}
After the diffusion is trained, we can generate the parameters $\bm \theta_{n, 0}^\text{a}$ by querying $g_{\bm \xi}$ with a new set of preference weights $\bm w_n$, specifying the desired preference weights for accuracy and diversity of new task $n$. Then the generated adapter parameter $\bm \theta_{n, 0}^\text{a}$ for that new task is directly loaded into the adapter, which is connected to the backbone. This forms a new customized recommendation model that responds to the preference weights of the new task.
The generation is an iterative sampling process from step $t=T$ to $t=0$, which denoises the Gaussian noise into meaningful parameters taking specific preference weights as the condition. The generation process is formulated as follows:
\begin{equation}
\begin{aligned}
    \tilde{\epsilon}_{\bm \xi}(\bm \theta_{n,t}^\text{a}, \bm w_n, t) &= (1+\gamma)\epsilon_{\bm \xi}(\bm{\theta}_{n,t}^\text{a},\bm w_n,t) - \gamma\epsilon_{\bm \xi}(\bm{\theta}_{n,t}^\text{a},t), \\
    \bm{\theta}_{n,t-1}^\text{a} &= \frac{1}{\sqrt{\alpha_t}}
    \left[ \bm{\theta}_{n,t}^\text{a} - \frac{\beta_t}{\sqrt{1-\overline{\alpha}_t}} \tilde{\epsilon}_{\bm \xi}(\bm \theta_{n,t}^\text{a}, \bm w_n, t) \right] + \sigma_t\bm{z}_t,
\end{aligned}
\end{equation}
\noindent where $\bm{z}_t\sim \mathcal{N}(\bm{0},\bm{I})$ for $t>1$ and $\bm{z}_t=\bm{0}$ for $t=1$, $\beta_t = 1-\alpha_t$, $\gamma \in [0,1]$ . 

Specifically, after generating the adapter parameter matrix, we reshape it to obtain the adapter parameters (for simplicity, we do not distinguish between the notations used before and after the reshaping). The generated adapter parameter is directly load into the adapter architecture. Then keeping the backbone parameters $\bm \theta_n^\text{b}$ and the adapter parameters $\bm \theta_{n,0}^\text{a}$ fixed, the recommendation model is directly applied to extract features from the user history interactions and score candidate items to generate a recommendation list that aligns with the preference weights of the new task.

\begin{table}[htbp]
\small
\centering
\caption{Statistical information of Datasets}
\resizebox{0.48\textwidth}{!}{
\begin{tabular}{lccc c c}
\toprule
\textbf{Dataset} & \textbf{\#user} & \textbf{\#item} & \textbf{\#category} & \textbf{\#inter.} & \textbf{density} \\
\hline
\textbf{\texttt{MovieLens 1M}}& 6,034 & 3,125 & 18 & 994,338 & 7.805\% \\
\textbf{\texttt{Amazon Food}} & 4,905 & 2,420 & 156 & 53,258 & 0.448\% \\
\textbf{\texttt{Industrial Data}} & 3,628 & 3,181 & 27 & 72,391 & 0.627\% \\
\bottomrule
\end{tabular}
}
\label{tab:dataset}
\end{table}

\section{Experiments}

\label{sec:diff:Experiments}

We conducted experiments to evaluate the performance of Paragon.
\subsection{Experiment Settings}

\begin{table*}[t]

\vspace{-0.3cm}
\caption{Performance comparison between the proposed method and baseline models. The \textbf{best} results are highlighted in bold, while the \underline{second-best} results are underlined. ``/'' represents the absence of a relevant value. }



\resizebox{1\textwidth}{!}{
\begin{tabular}{l|l|c|c|c|c|c|c|c|c|c}
\toprule[1pt]
& &
\multicolumn{3}{c|}  {\textbf{\texttt{MovieLens}}} & \multicolumn{3}{c|}{\textbf{\texttt{Amazon Food}}}&\multicolumn{3}{c}{\textbf{\texttt{Industrial Dataset}}} \\
 \cline{3-11} 

\multirow{-2}{*}{\textbf{Backbone}} & \multirow{-2}{*}{\textbf{Algorithm}} & Avg.HV & pearson r-a& pearson r-d& Avg.HV & pearson r-a  & pearson r-d & Avg.HV & pearson r-a& pearson r-d  \\
\midrule 

&Retrain & \textbf{0.2281} & /&/ & \underline{0.2251} &/& / & \underline{0.2779} &/& / \\

 &CMR& 0.1920 & 0.8901& \underline{0.9150} & 0.1955 & -0.7039 & \textbf{0.9932}& 0.2476 &0.8750 & 0.9237 \\
 
 & Soup & 0.1441 & 0.7861 &  0.9133 &0.1561 & 0.5317& 0.6693& 0.1825 & 0.7306& 0.8188 \\
 
 & MMR & 0.1808 &\underline{0.9575}&  0.8803 &0.1707 &0.1320 & -0.3087 & 0.2034 &\underline{0.9077}& \underline{0.9655} \\
 
\multirow{-5}{*}{SASRec} & \cellcolor[HTML]{F2F2F2}\textbf{Paragon}  &\cellcolor[HTML]{F2F2F2} \underline{0.2138} & \cellcolor[HTML]{F2F2F2}\textbf{0.9905}& \cellcolor[HTML]{F2F2F2}\textbf{0.9903} & \cellcolor[HTML]{F2F2F2}\textbf{0.2420} & \cellcolor[HTML]{F2F2F2}\textbf{0.8857} & \cellcolor[HTML]{F2F2F2}\underline{0.9816}  & \cellcolor[HTML]{F2F2F2}\textbf{0.2812} & \cellcolor[HTML]{F2F2F2}\textbf{0.9976}& \cellcolor[HTML]{F2F2F2}\textbf{0.9986}  \\
\cline{1-1}

/&LLM-CMTR &  0.0625 &   -0.0600&  0.0994 &0.1017 &  \underline{0.7296} & 0.8558& 0.0372 &  0.7279 & 0.7132 \\
\midrule

 &Retrain & \underline{0.1823} & /& / & 0.1556& /& / & \underline{0.1735} & /& / \\
 
 &CMR& 0.1760 & \underline{0.9068}  & \underline{0.8813} & \textbf{0.3617} & \underline{0.8287}  &  \underline{0.9059} & 0.1230 & 0.6514 & 0.5985 \\

 & Soup & 0.1197 & 0.8061& 0.6694 &0.0604 & 0.3850& 0.8005 & 0.1226 & 0.7099& \underline{0.8200} \\
 
 & MMR & 0.1609 & 0.8692& 0.7257 &0.1354 & -0.3497& -0.3748& 0.1287 & 0.7916& 0.7553 \\
 
\multirow{-5}{*}{GRU4Rec} & \cellcolor[HTML]{F2F2F2}\textbf{Paragon}  & \cellcolor[HTML]{F2F2F2}\textbf{0.2009} & \cellcolor[HTML]{F2F2F2}\textbf{0.9929}& \cellcolor[HTML]{F2F2F2}\textbf{0.9786} &\cellcolor[HTML]{F2F2F2} \underline{0.1623} &\cellcolor[HTML]{F2F2F2} \textbf{0.8470}&\cellcolor[HTML]{F2F2F2} \textbf{0.9685}  &\cellcolor[HTML]{F2F2F2} \textbf{0.1871} & \cellcolor[HTML]{F2F2F2}\textbf{0.9760}&\cellcolor[HTML]{F2F2F2} \textbf{0.9236}  \\
\cline{1-1}
/&LLM-CMTR & 0.0625 & -0.0716 & 0.0119 &0.0667 & 0.7484&  0.7904& 0.0372 & \underline{0.8088}& 0.7722 \\

\midrule

&Retrain & \underline{0.2301} & /& / & \underline{0.2232}& /& / & 0.2777 & /& / \\

 &CMR& 0.1769 &  \underline{0.9286} & \underline{0.9903}  & 0.2064 & -0.6279 & \underline{0.9828} & \textbf{0.3315} & 0.8855& 0.9300  \\
 
 & Soup & 0.1483 & 0.8033& 0.8858&0.1533& 0.5342& 0.6525 & 0.1811 & 0.7310& 0.8248 \\
 
 & MMR & 0.1815 & 0.8946& 0.8684 &0.1672 & 0.3057& 0.2037 & 0.2060 & \underline{0.9004}& \underline{0.9565} \\
 
\multirow{-5}{*}{TiSASRec} & \cellcolor[HTML]{F2F2F2}\textbf{Paragon}  & \cellcolor[HTML]{F2F2F2}\textbf{0.2532} & \cellcolor[HTML]{F2F2F2}\textbf{0.9923}&  \cellcolor[HTML]{F2F2F2}\textbf{0.9914}& \cellcolor[HTML]{F2F2F2}\textbf{0.2394} & \cellcolor[HTML]{F2F2F2}\textbf{0.8759}& \cellcolor[HTML]{F2F2F2}\textbf{0.9851}  & \cellcolor[HTML]{F2F2F2}\underline{0.2862} &\cellcolor[HTML]{F2F2F2} \textbf{0.9968}& \cellcolor[HTML]{F2F2F2}\textbf{0.9984}  \\
\cline{1-1}
/&LLM-CMTR &  0.0625 & -0.0663& 0.0999 &0.0667 & \underline{0.7213}& 0.8499& 0.0372 & 0.7373& 0.7451 \\

\bottomrule[1pt]
\end{tabular}
}

\vspace{-0.4cm}

\label{tab:main}
\end{table*}

\subsubsection{Dataset}

The datasets are processed as follows:

\textbf{MovieLens-1M}~\footnote{https://grouplens.org/datasets/movielens/} is an website dataset of MovieLens in 2000. We sorted each user's browsing history chronologically and filtered out users with fewer than 5 interactions. Each interaction is formatted to include user ID, item ID, timestamp, Categories of items (may be multiple categories).  

\textbf{Amazon Grocery and Gourmet Food}~\footnote{http://jmcauley.ucsd.edu/data/amazon/links.html} is the food data from Amazon website, spanning from August 09, 2000 to July 23, 2014. Since the items belong to 156 categories, we used the GloVe~\citep{pennington2014glove} to generate embeddings for each category. We then applied K-means clustering to group them into 30 broader categories. The interaction format is the same as MovieLens 1M.

\textbf{The industrial dataset} is the user click dataset from an electronics commercial store, spanning from July 24, 2024, to August 24, 2024. We randomly sample 10000 user, then filtering out users with fewer than 10 interactions to obtain 3628 users. Each interaction was formatted to match the structure used in MovieLens 1M.  The specific statistical information of the three datasets
is in Table~\ref{tab:dataset}.

\subsubsection{Baselines} 
\label{sec:appendix:baselines}
To validate its effectiveness, we compare our model against several baseline methods adapted for CMTR.

\textbf{Retraining} is performed using Linear Scalarization~\citep{birge2011introduction} based on task requirements, representing the optimal solution without considering the associated costs.

\textbf{Soup}~\citep{wortsman2022model} trains separate models for each aspect of the task requirement and merges them linearly during the testing stage.

\textbf{MMR}~\citep{carbonell1998use} is a heuristic post-processing approach with the item selected sequentially according to maximal marginal relevance. 

\textbf{CMR}~\citep{chen2023controllable} dynamically adjusts models based on preference weights using policy hypernetworks to generate model parameters.

\textbf{LLM-CMTR}~\citep{dai2023uncovering} is a prompt-based method but specifically customized for CMTR. It inputs prompts containing specific preference weights to guide the LLM's generation in the form of list-wise recommendations. In our experiments, we selected the llama3-7B-Instruct model. 

\subsubsection{Implement Details}
  We first filtered for 5-core user IDs and adopted a leave-one-out data splitting strategy to divide the dataset into training, validation, and test sets. Additionally, we limited the length of the target item's interaction history to no more than 20. For the three recommendation model backbones, we follow the settings used in RecChorus, where the Adam optimizer is applied with a learning rate of 1e-3, embedding size of 64, and hidden size of 64. The number of negative samples in the training set is set to 9, and 99 for both the validation and test sets.  

\subsubsection{Metrics} 
We evaluate the algorithms from two dimensions. Specifically, we use Hypervolume (HV)~\citep{guerreiro2021hypervolume} to measure the performance of the algorithm on each task, particularly in terms of the trade-offs between accuracy and diversity. The average HV (denoted as \textbf{Avg.HV}) across multiple tasks is used to assess the overall performance of the algorithm in balancing both objectives (accuracy and diversity). To eliminate the differences in scale between the two objectives, we normalize the performance on each objective. Additionally, we utilize the Pearson correlation coefficient to evaluate the alignment between the algorithm's performance across different tasks and the optimal model, providing insight into the algorithm's controllability. \textbf{Pearson r-a}, \textbf{Pearson r-d} measure the correlation between the algorithm and the optimal in terms of accuracy and diversity. Notably, the accuracy and diversity is measured by NDCG$@10$ and $\alpha$-NDCG@10). The fairness in analysis experiment is measured by Absolute Difference (AD)~\citep{wang2023survey}.

\begin{figure}[t]
    \centering
    \begin{subfigure}[b]{0.22\textwidth}  
        \centering
       \includegraphics[width=\textwidth]{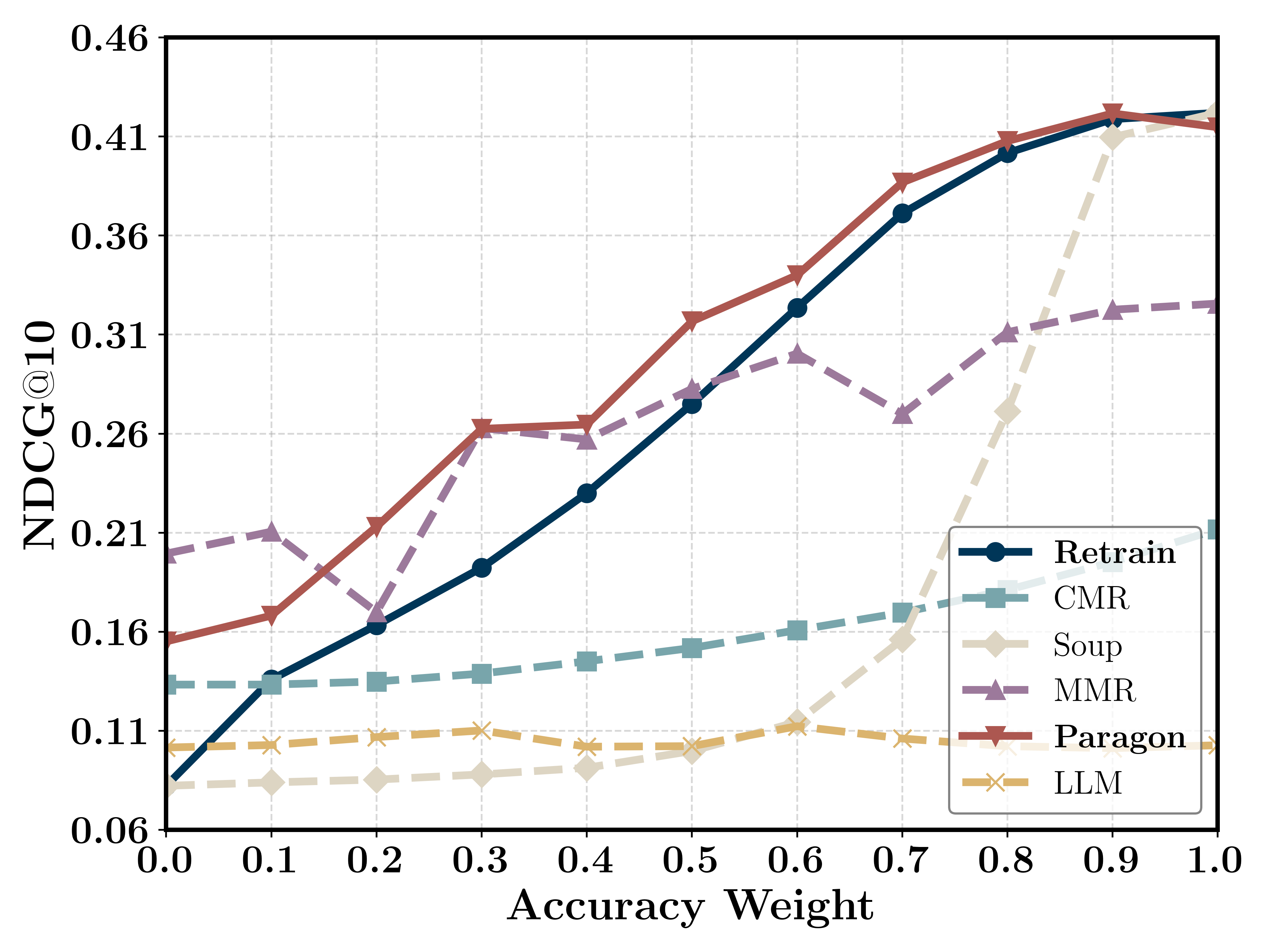} 
        \caption{Accuracy on \texttt{MovieLens 1M}}
    \end{subfigure}
    \hfill
    \begin{subfigure}[b]{0.22\textwidth}
        \centering
        \includegraphics[width=\textwidth]{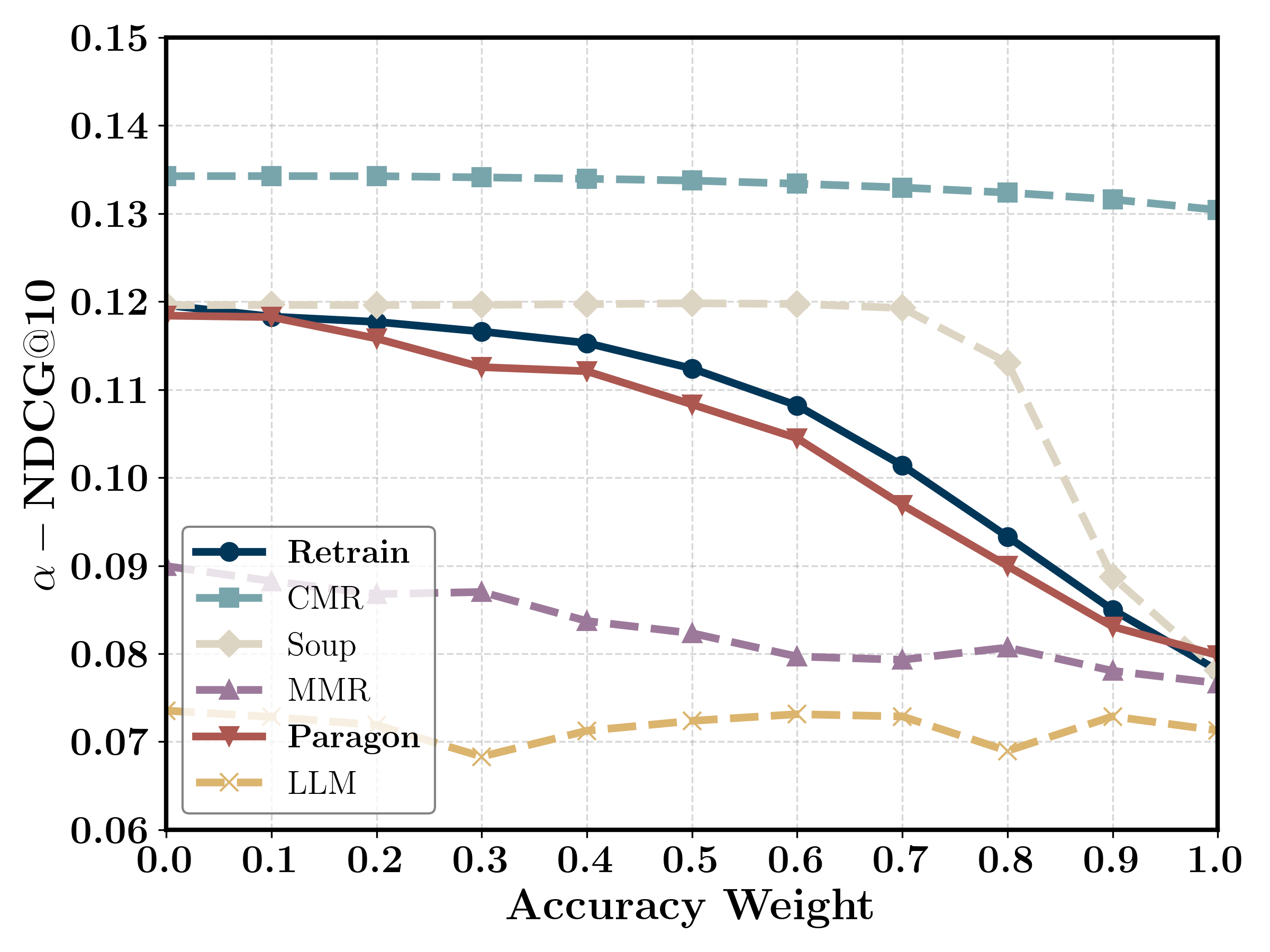}
        \caption{Diversity  on \texttt{MovieLens 1M}}
    \end{subfigure}
    
    \begin{subfigure}[b]{0.22\textwidth}  
        \centering
        \includegraphics[width=\textwidth]{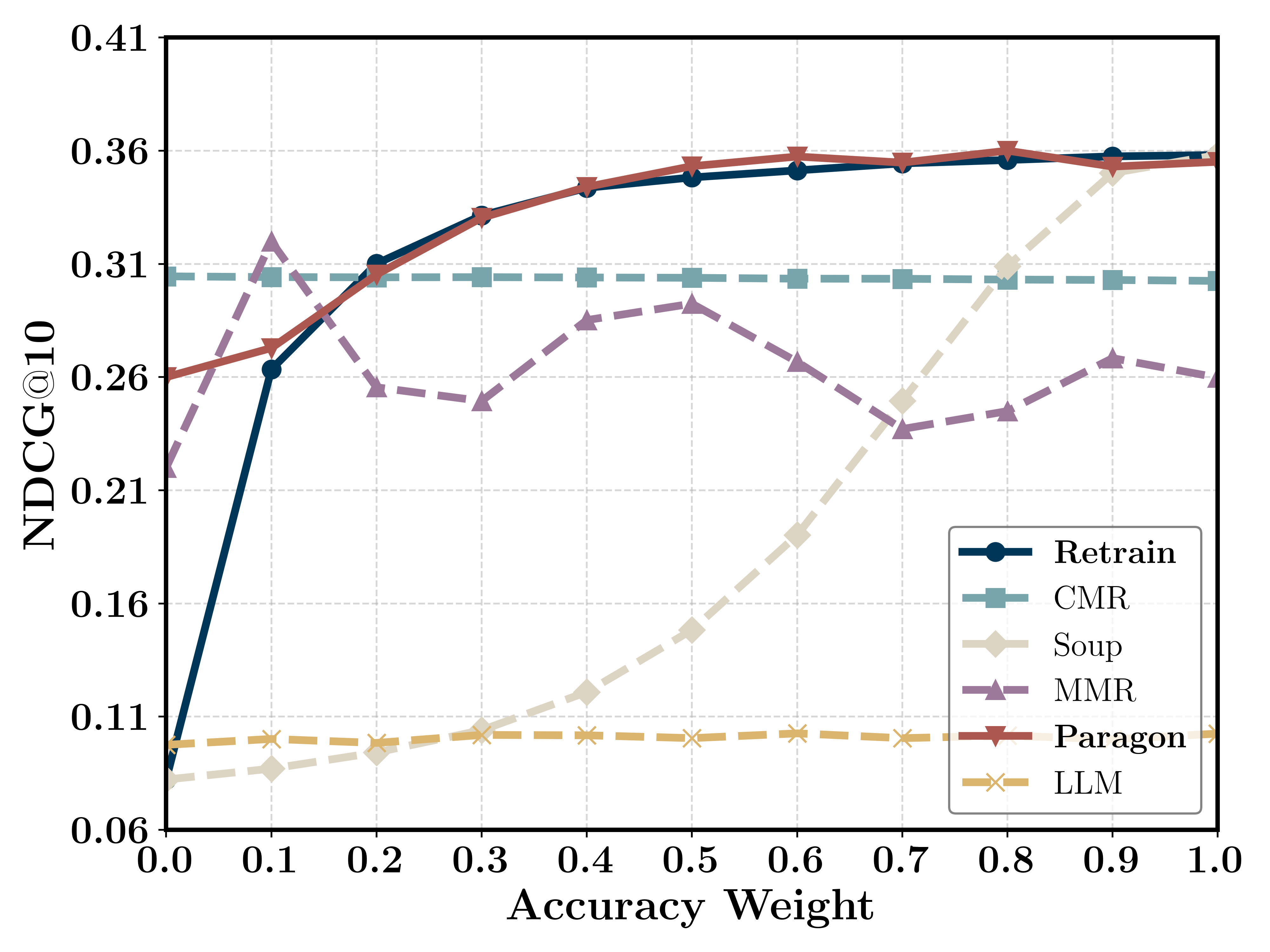}
        \caption{Accuracy on \texttt{Amazon Food}}
    \end{subfigure}
     \hfill
    \begin{subfigure}[b]{0.22\textwidth}
        \centering
        \includegraphics[width=\textwidth]{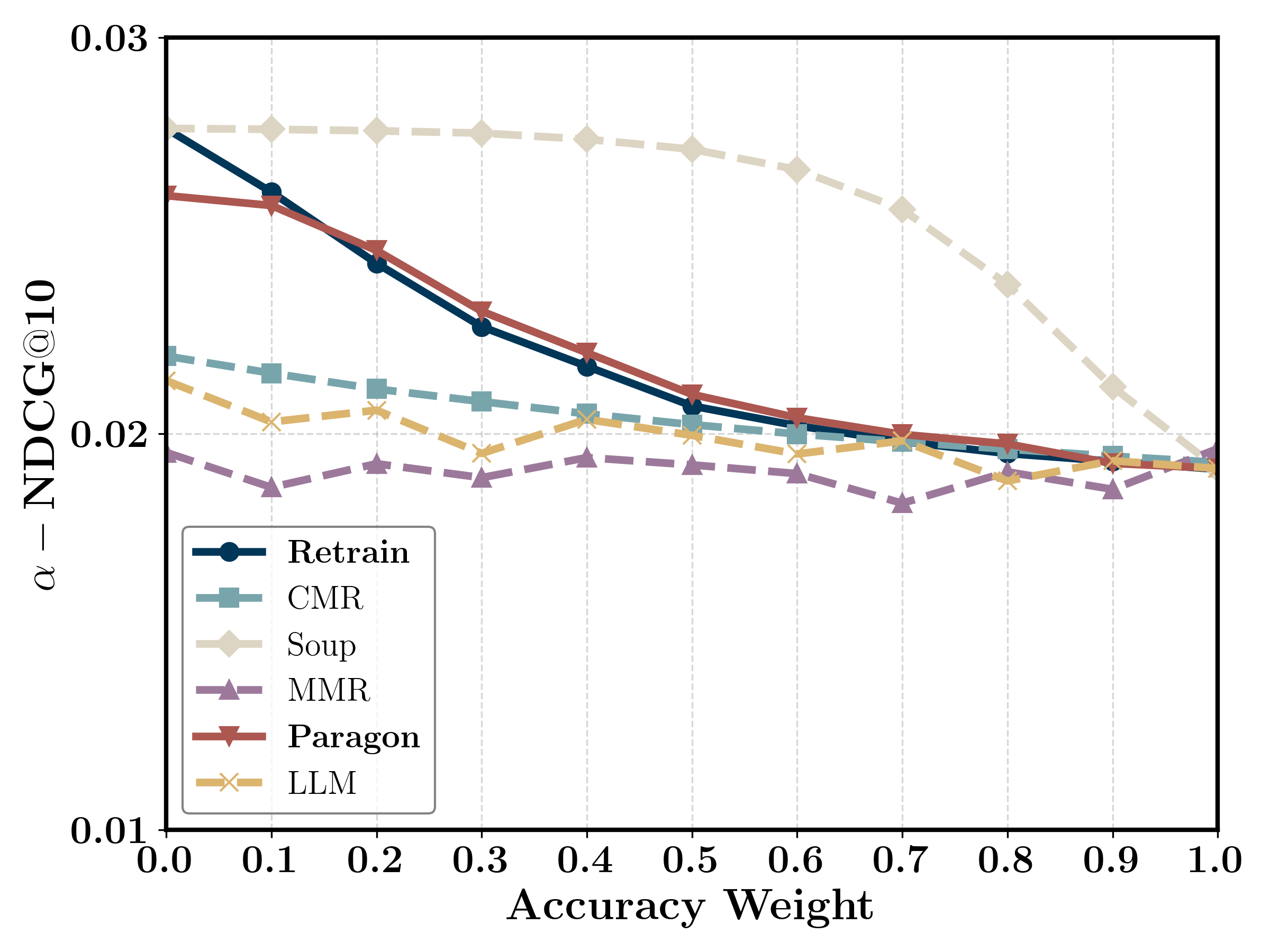}
        \caption{Diversity on \texttt{Amazon Food}}  
    \end{subfigure}
    
    \begin{subfigure}[b]{0.22\textwidth}
        \centering
        \includegraphics[width=\textwidth]{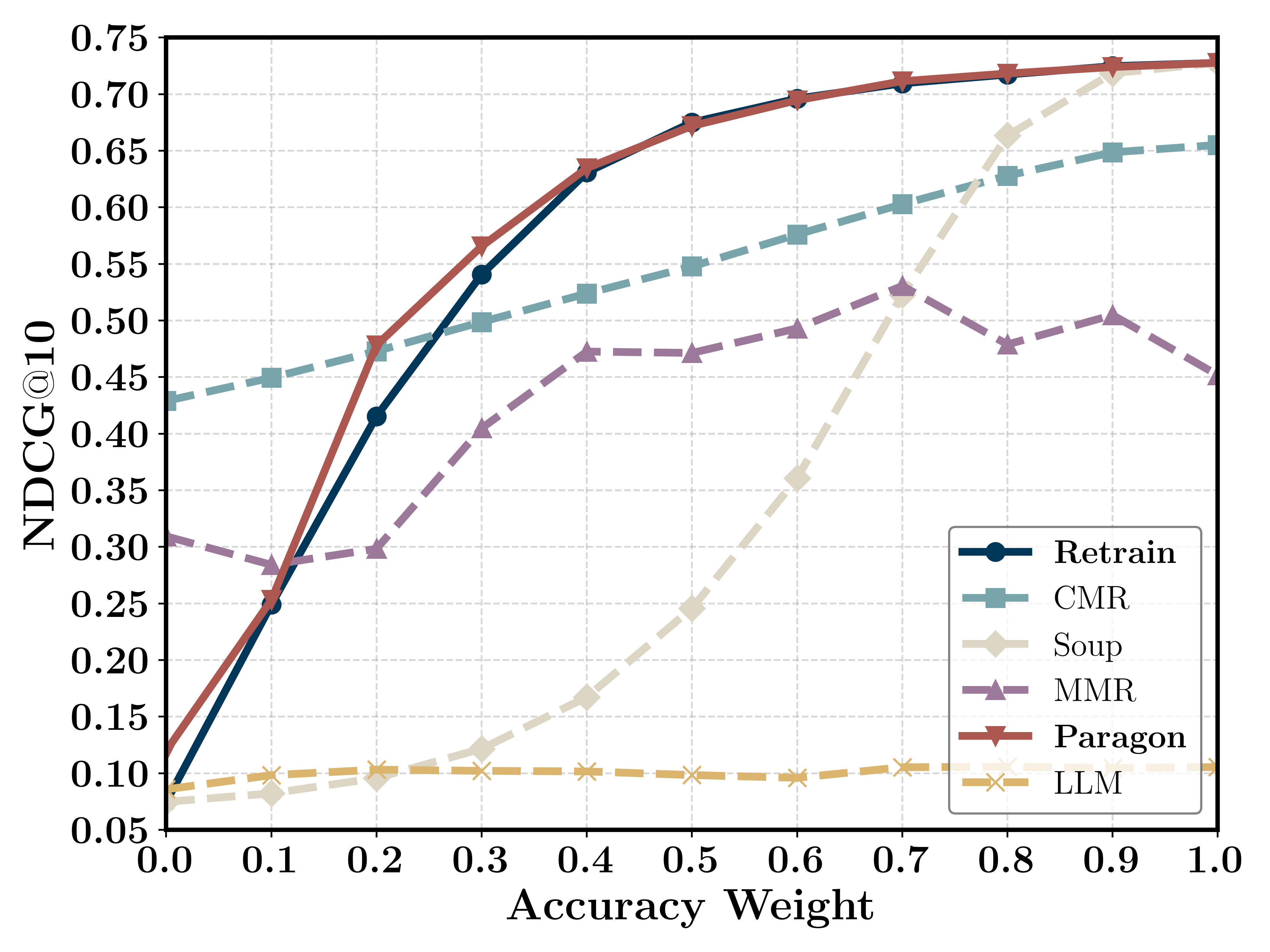}
        \caption{Accuracy on Industrial Data}
    \end{subfigure}
    \hfill
    \begin{subfigure}[b]{0.22\textwidth}
        \centering
        \includegraphics[width=\textwidth]{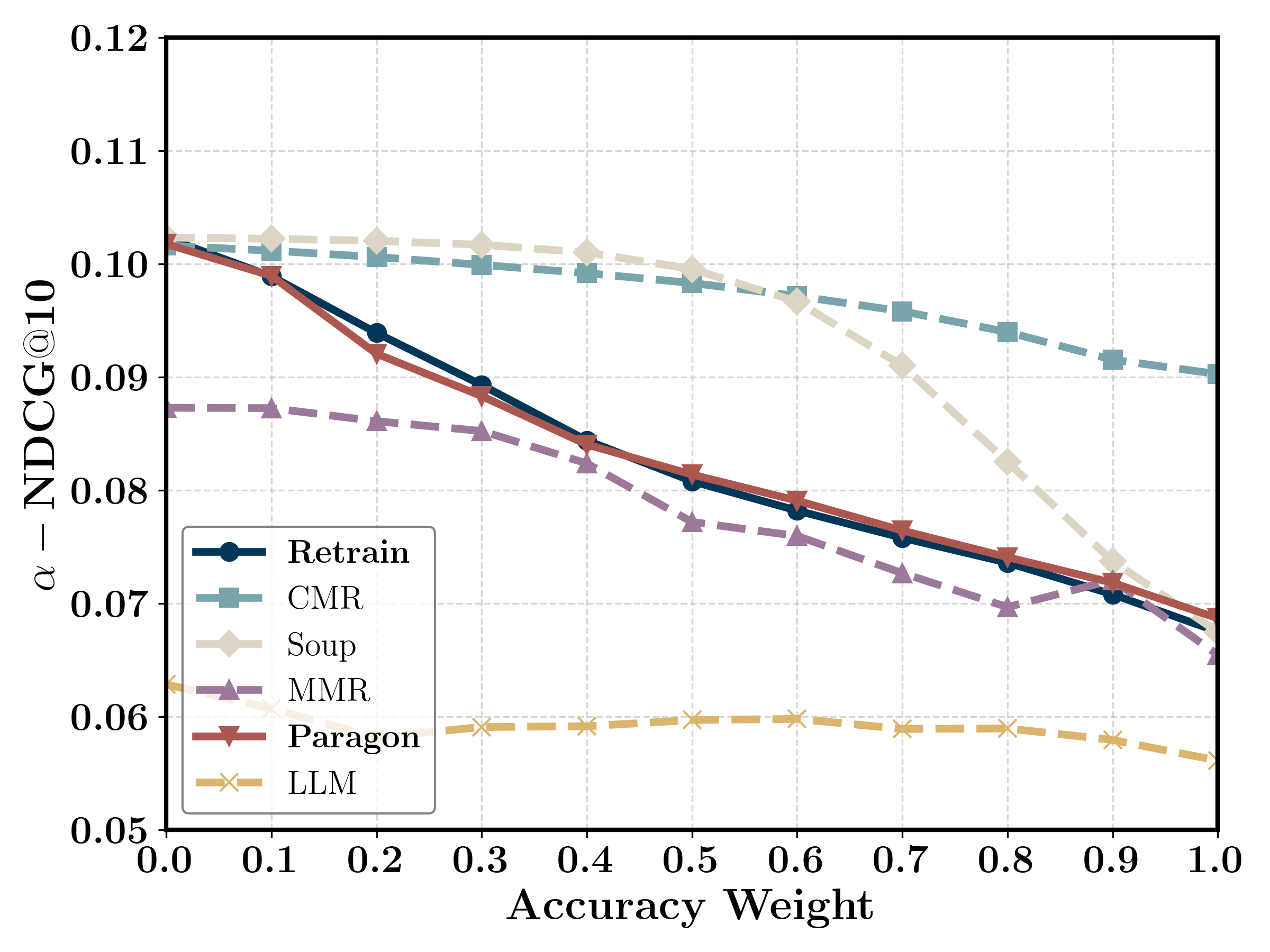}
        \caption{Diversity on Industrial Data}
    \end{subfigure}

    \caption{The accuracy and diversity curve of Paragon and other baselines in NDCG@10 and $\alpha$-NDCG@10 across accuracy weights ranging from 0 to 1, with intervals of 0.1. The backbone is TiSASRec.}
    \label{fig:tend:TiSASRec}
    
\end{figure}

\subsection{Experimental Results}

We conducted experiments to address the following two questions: i) How transferable is Paragon, specifically in terms of its ability to adapt to different backbone algorithms? ii) How does Paragon perform compared to other baselines? The results are presented in Table~\ref{tab:main}.

To answer the first question, we used commonly adopted sequential recommendation models as backbones (e.g., SASRec~\citep{kang2018self}, GRU4Rec~\citep{hidasi2015session}, and TiSASRec~\citep{li2020time}) and conducted extensive experiments across three datasets. 
Specifically, we evaluated Paragon’s performance under various task descriptions by measuring NDCG@10 and $\alpha$-NDCG@10. The accuracy weight $w_{\text{acc.}}$ varies from 0 to 1 in intervals of 0.1, with the corresponding diversity weight set as $w_{\text{div.}}=1 - w_{\text{acc.}} $.  We then post-processed NDCG@10 and $\alpha$-NDCG@10 across different tasks to compute Avg.HV, Pearson r-a, and Pearson r-d. These metrics respectively evaluate the quality of multi-objective optimization on individual tasks and the controllability across multiple tasks.

Overall, across the three backbones and three datasets, Paragon consistently ranked among the top two performers across all three evaluation metrics. Notably, in most cases, the top two of Avg.HV are Retrain and Paragon, indicating that Paragon’s performance in multi-objective trade-offs is on par with, or even superior to, Retrain. Specific exceptions occurred, such as on the \texttt{Amazon Food} dataset with GRU4Rec as the backbone and the \texttt{Industrial Data} with TiSASRec as the backbone, where CMR achieved the best Avg.HV. This is because CMR is not influenced by task descriptions and thus maintains consistently high NDCG@10 scores (as shown in Figure~\ref{fig:tend:TiSASRec} and further explained in response to question ii).  For Pearson r-a and Pearson r-d, Paragon demonstrated strong correlations with the Retrain method, indicating that Paragon closely aligns with Retrain (which we assume to be optima) in terms of accuracy (NDCG@10) and diversity ($\alpha$-NDCG@10) across different tasks. On the \texttt{Amazon Food} dataset with SASRec as the backbone, CMR achieved the highest Pearson r-d. However, its Pearson r-a was negative, indicating a lack of control and a collapse in accuracy.
\begin{figure}[t]
    \centering
    \includegraphics[width=0.96\linewidth]{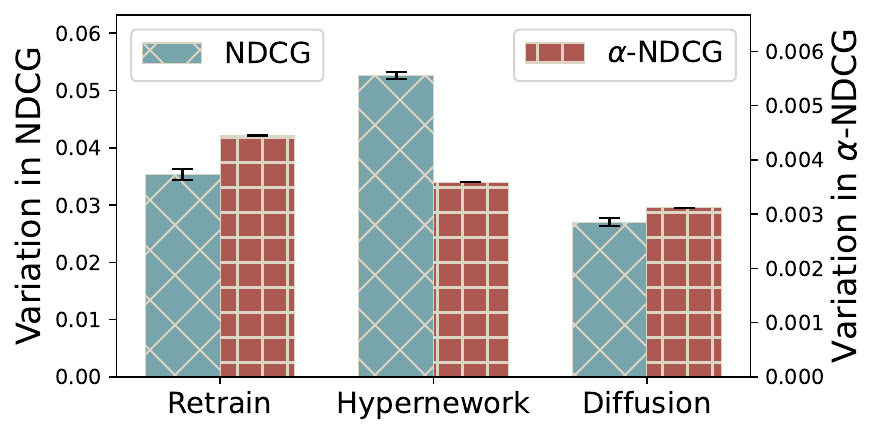}
    \caption{The variation of performance before and after disturbance in \texttt{MovieLens 1M} based on SASRec. The blue bars represent the variation in NDCG@10, while the red represent the variation in $\alpha$-NDCG@10 }
    \label{fig:disturbance}
\end{figure}
To address the second question, we presented the specific performance of Paragon under each task description using TiSASRec as the backbone across three datasets, as shown in Figure~\ref{fig:tend:TiSASRec}. It is observed that in all three datasets, Paragon's NDCG@10 progressively increases with higher accuracy weights, while $\alpha$-NDCG@10 decreases correspondingly due to the simultaneous reduction in diversity weight. These trends demonstrate the effectiveness of our algorithm in controllability. Notably, assuming that Retrain is optimal, Paragon exhibits strong consistency with the Retrain method. In contrast, MMR, as a post-processing algorithm, shows variability because varying degrees of diversity manipulation can disrupt the original recommendation list, uncontrollably affecting its accuracy. The Soup method merges the parameters of accuracy and diversity models based on their weights, aligning closely with the Retrain model when accuracy weights are extreme but showing significant deviations in other tasks. This indicates that tasks do not follow a simple linear relationship with different preference weights, and Soup makes overly strong assumptions about this relationship. CMR demonstrates inconsistent performance across different datasets. On the \texttt{MovieLens 1M} dataset, CMR aligns well with the original descriptions by exhibiting high diversity. However, on the other two datasets, it shows a stable yet uncontrollable state; for instance, on the \texttt{Amazon Food} dataset, CMR maintains high accuracy even with low accuracy weights, and on the \texttt{Industrial Data}, it retains high diversity despite low diversity weights.

\begin{figure*}[h]
    \centering
    \begin{subfigure}{0.45\textwidth}  
       \centering
       \includegraphics[width=\textwidth]{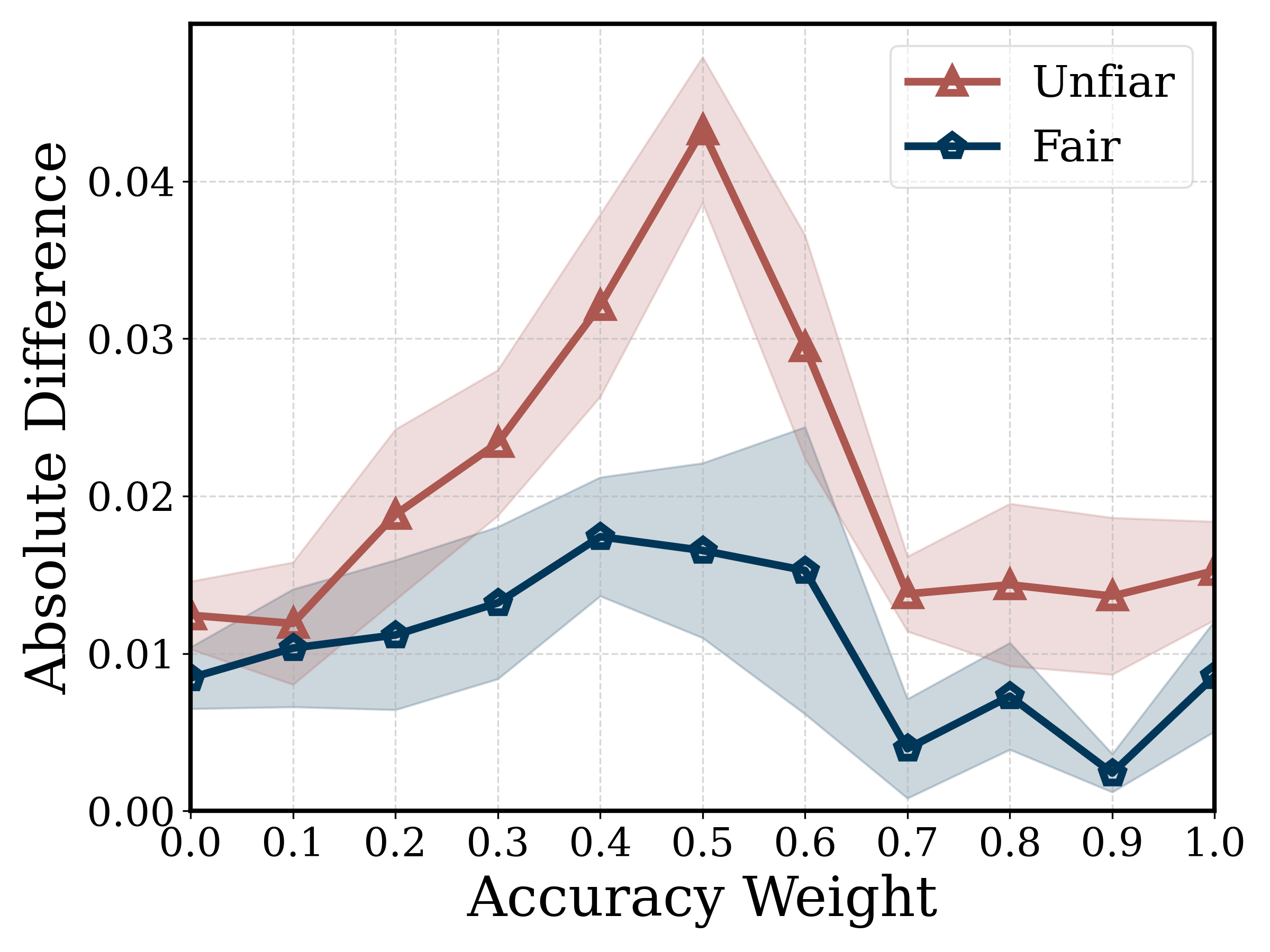} 
        \caption{The result of fairness.}
    \end{subfigure}
    \hfill
    \begin{subfigure}{0.45\textwidth}  
        \centering
        \includegraphics[width=\textwidth]{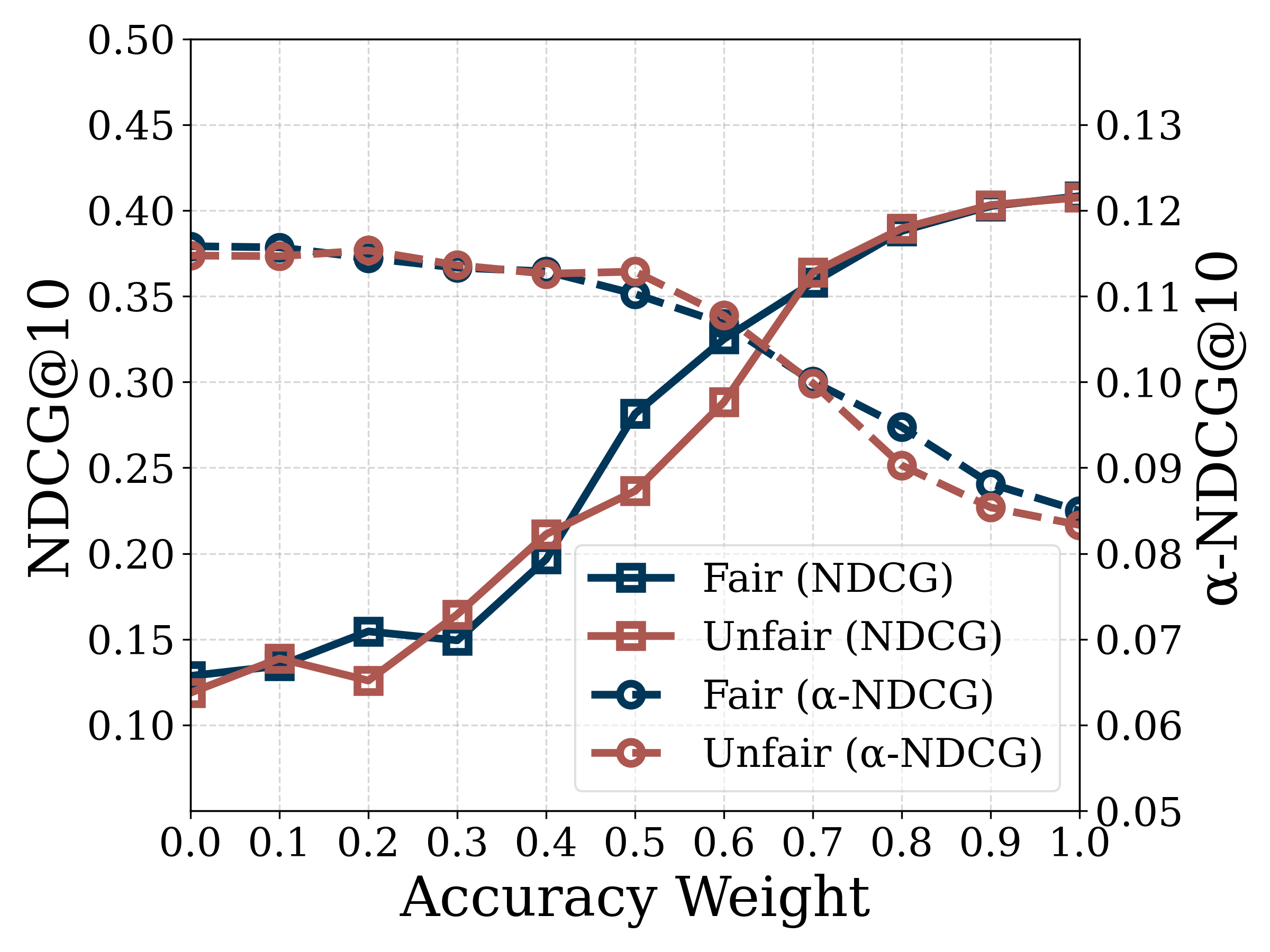}
        \caption{The result of acc. and div.}
    \end{subfigure}

    \caption{The result of control under accuracy, diversity and fairness. The accuracy weight ranges from 0 to 1, with intervals of 0.1. The diversity weight is set to $1-$ accuracy weight. The fairness weight is $\{1,0\}$ corresponding to `fair' and `unfair'. The experiment is conducted on the MovieLens 1M, uti-lizing SASRec as the backbone.}
    \label{fig:fair}
\end{figure*}

\subsection{Analyses}
\label{sec:analysis}
We conducted our analysis  experiments based on three key research questions:

\begin{table}[h]
\centering
\normalsize
\renewcommand{\arraystretch}{1.5}
\arrayrulewidth=0.5pt 
\caption{Response time comparison between proposed Paragon and the ``Retrain'' approach across three datasets using three different backbones. Note that the unit is seconds (sec.).  }
\resizebox{0.48\textwidth}{!}{

\begin{tabular}{c|c|c|c|c}

\toprule
\textbf{Approach} & \textbf{Backbone} & \texttt{MovieLens 1M} (sec.) & \texttt{Amazon Food} (sec.) & \texttt{Industrial Data} (sec.)\\ \hline
\multirow{3}{*}{Retrain} & SASRec & 293.10 ± 11.61 & 91.01 ± 2.34 & 46.82 ± 3.25 \\ 
 \cline{2-5}
 
 & GRU4Rec & 281.60 ± 17.36 & 92.39 ± 4.28 & 49.54 ± 2.38 \\ 
 \cline{2-5}
 
 & TiSASRec & 303.80 ± 9.09 & 105.40 ± 7.66 & 52.47 ± 4.64 \\ 
 \midrule
\multirow{3}{*}{Paragon} & SASRec & 2.68 ± 0.36\textsuperscript{\textcolor{red}{\scriptsize{-99.1\%}}} & 2.64 ± 0.36\textsuperscript{\textcolor{red}{\scriptsize{-97.1\%}}} & 2.55 ± 0.25\textsuperscript{\textcolor{red}{\scriptsize{-94.6\%}}} \\ \cline{2-5} 

 & GRU4Rec & 2.56 ± 0.27\textsuperscript{\textcolor{red}{\scriptsize{-99.1\%}}}  & 2.54 ± 0.24\textsuperscript{\textcolor{red}{\scriptsize{-97.3\%}}}  & 2.51 ± 0.23 \textsuperscript{\textcolor{red}{\scriptsize{-94.9\%}}} \\ 
 \cline{2-5}
 
 & TiSASRec & 2.55 ± 0.23\textsuperscript{\textcolor{red}{\scriptsize{-99.2\%}}}  & 2.52 ± 0.24\textsuperscript{\textcolor{red}{\scriptsize{-97.6\%}}}  & 2.58 ± 0.26 \textsuperscript{\textcolor{red}{\scriptsize{-95.1\%}}} \\
\bottomrule

\end{tabular}
}

\label{tab:time}
\end{table}

\subsubsection{\textbf{RQ1: What are the advantages of Diffusion over Hypernetwork in parameter generation? }}

We conducted experiments to validate the robustness of the parameters generated by Paragon. Specifically, we designed three sets of experiments to constructed adapter parameters: ``Retrain'', ``Diffusion'' and ``Hypernetwork'', where ``Hypernetwork''  utilizes the MLP to learn the relationship between the preference weight and the optimized adapter parameters. First, we add Gaussian noise of the same magnitude to all three sets of adapter parameters and measured the resulting fluctuations in NDCG@10 and $\alpha$-NDCG@10.  We observed that the parameters generated by Paragon exhibited the lowest performance fluctuations than others, both in terms of accuracy (NDCG@10) and diversity ($\alpha$-NDCG@10) as depicted in Figure~\ref{fig:disturbance}. We could draw the following conclusions: 1) the Diffusion approach demonstrates superior robustness in parameter generation across both accuracy and diversity metrics compared to alternative methodologies. Second, Hypernetwork-based methods exhibit inferior robustness in parameter generation, a phenomenon likely attributable to the inherent limitations of discriminative models in achieving optimal generative capacity.
\begin{figure*}[t]
    \centering
    \begin{subfigure}{0.15\textwidth}  
        \centering
       \includegraphics[width=1.05\textwidth]{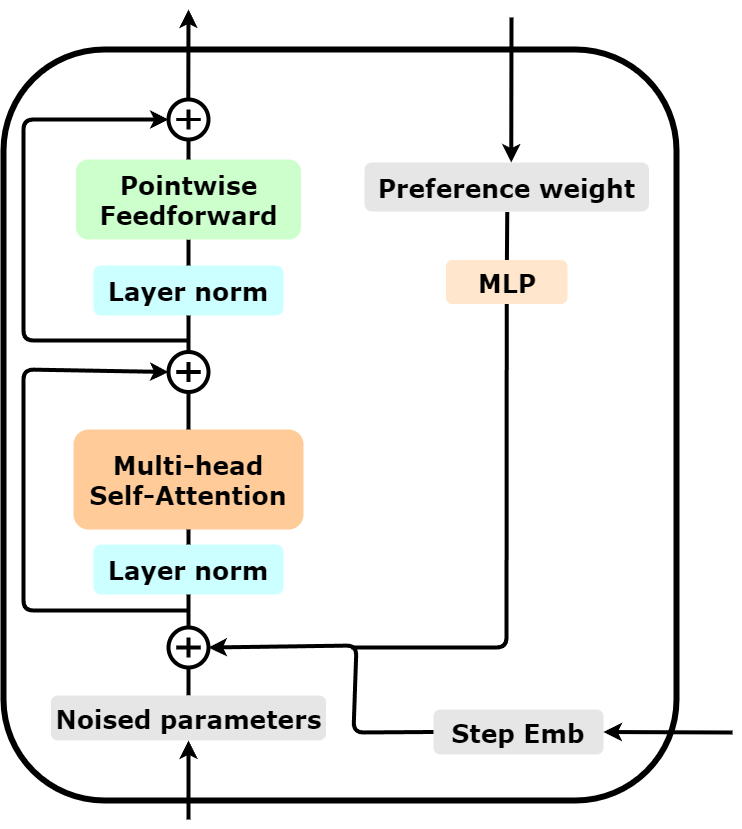} 
        \caption{Pre.}
    \end{subfigure}
    \hfill
    \begin{subfigure}{0.15\textwidth}  
        \centering
        \includegraphics[width=1.06\textwidth]{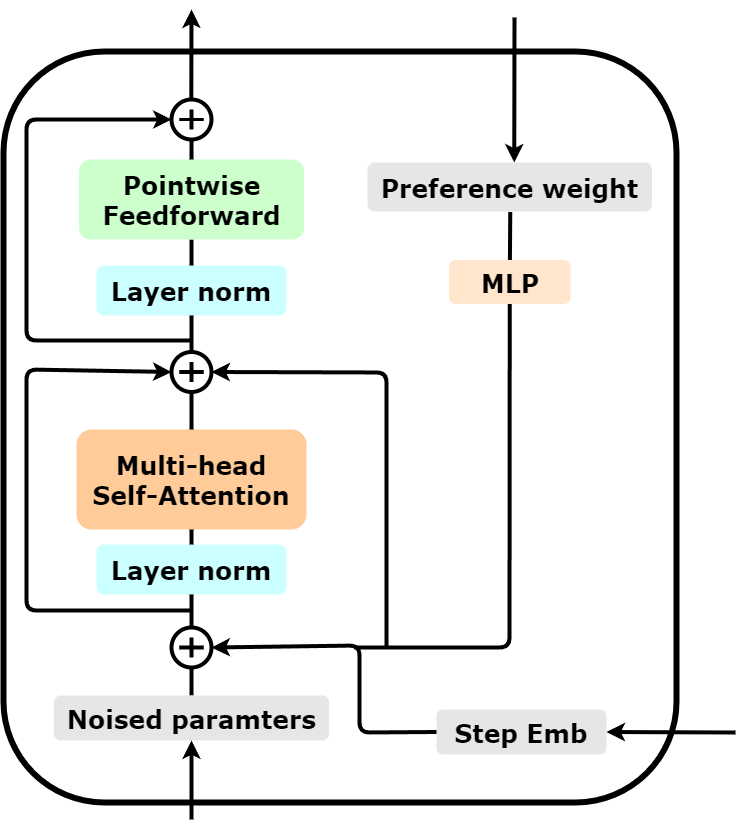}
        \caption{Pre.\&Post.}
    \end{subfigure}
    \hfill
    \begin{subfigure}{0.15\textwidth}
        \centering
        \includegraphics[width=\textwidth]{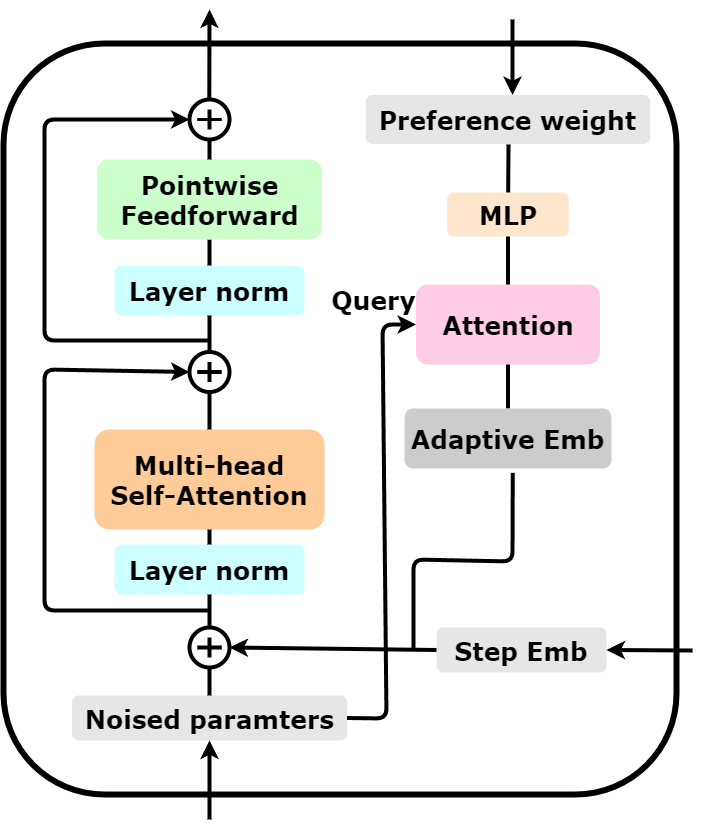}
        \caption{Pre-Adaptive}
    \end{subfigure}
    \hfill
    \begin{subfigure}{0.15\textwidth}
        \centering
        \includegraphics[width=\textwidth]{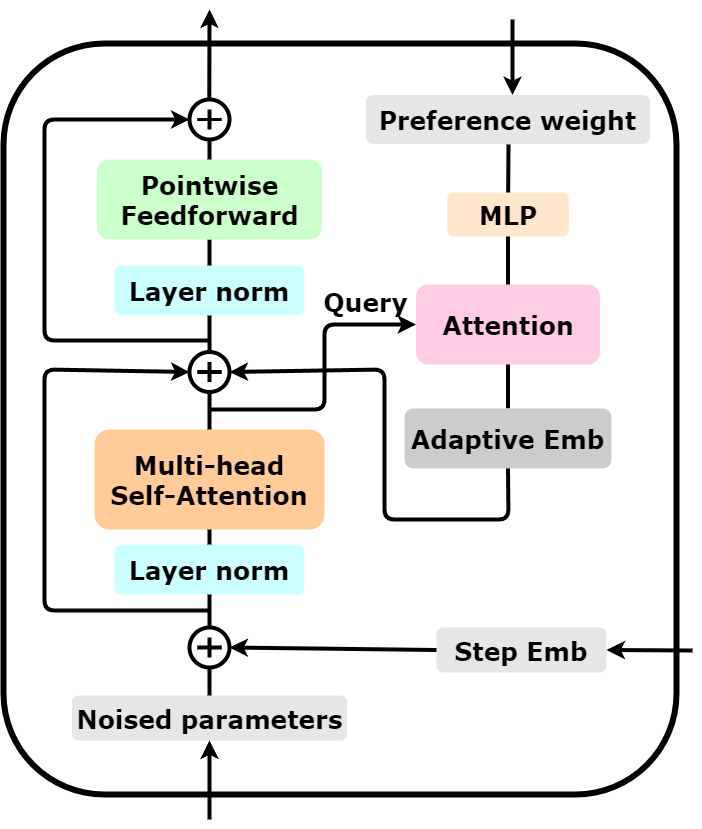}
        \caption{Post-
Adaptive}
    \end{subfigure}
    \hfill
    \begin{subfigure}{0.15\textwidth}
        \centering
        \includegraphics[width=\textwidth]{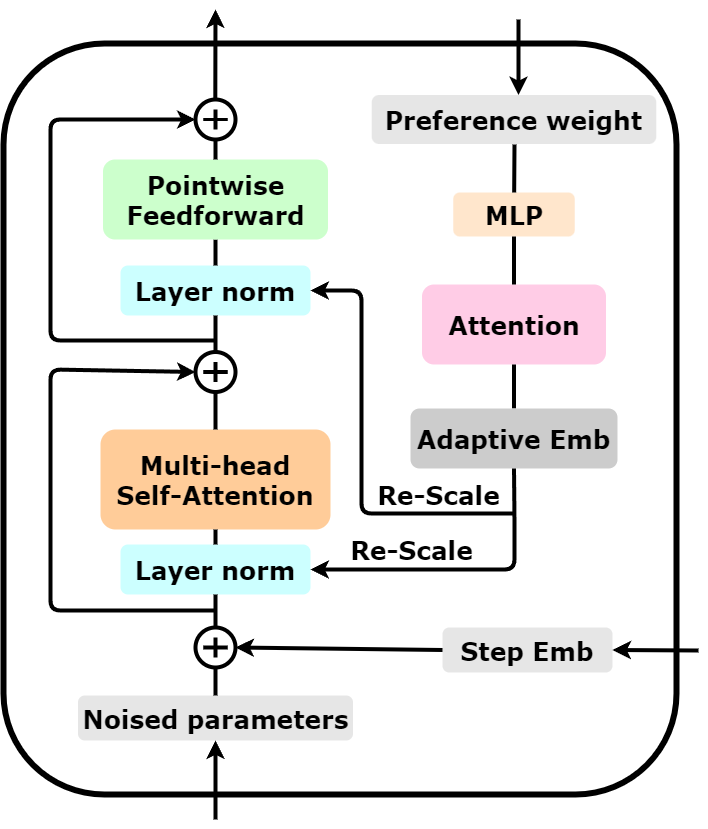}
        \caption{Adaptive-
Norm}
    
    \end{subfigure}
    \caption{Illustration of the structural diagram for five conditioning strategies.}
    \label{fig:prompt}
\end{figure*}

\subsubsection{\textbf{RQ2: Is Paragon efficient enough to handle real-time changes in preference weights compared to Retrain?} }
\begin{figure}[tb]
    \centering
    \includegraphics[width=\linewidth]{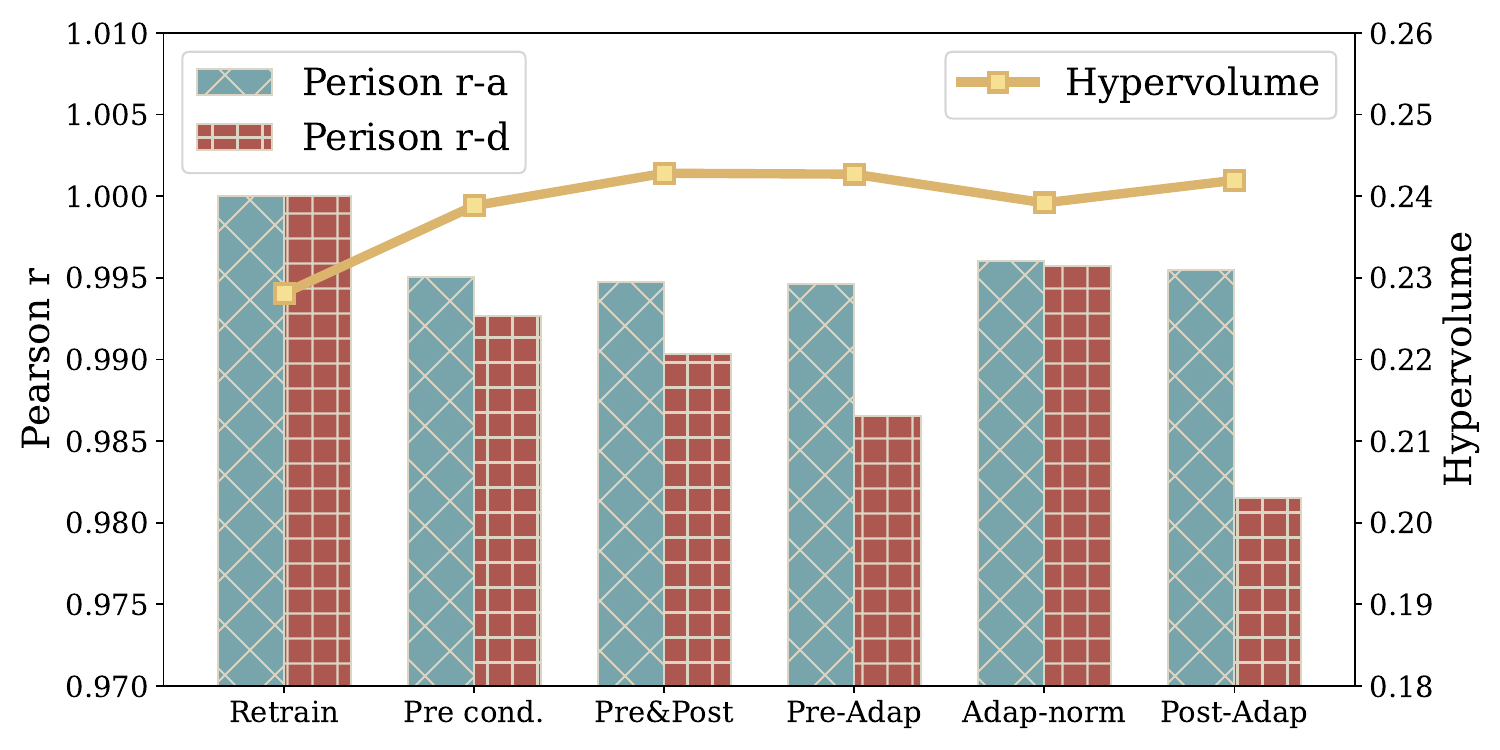}
    \caption{Performances of different conditioning strategies on \texttt{MovieLens 1M} using SASRec as backbone. The results of the ``Retrain'' algorithm are used as a reference. }

    \label{fig:condition_result}
\end{figure}
Paragon is designed to adaptively adjust model parameters in an online environment without retraining, enabling it to quickly respond to new task requirements. This places a strong emphasis on the model's response time. We compared the response times of Paragon and ``Retrain'' across various backbones and datasets, with the results shown in Table~\ref{tab:time}. As observed, in all experiments using three different backbones across three datasets, Paragon’s response time was significantly faster than that of  ``Retrain''. Notably, the response time of ``Retrain'' correlated with the size of the dataset, whereas Paragon exhibited minimal variation across different datasets. This highlights Paragon’s data-agnostic nature indicating its potential for handling large-scale datasets efficiently.

\subsubsection{\textbf{RQ3: What about the  scalability to more utilities such as fairness. }}

We examined the performance of Paragon under various metric controls, including the widely used fairness metric in recommendation systems, where fairness is defined as the absolute difference (AD) of NDCG@10 between male and female groups. We set the value of fairness dimension in preference weight to $\{0, 1\}$, corresponding to unfair and fair settings. As shown in Figure~\ref{fig:fair} (a), under different levels of control over accuracy and diversity, there is a clear and consistent difference in the models' AD. when set control signal to fair versus unfair. A smaller AD for the fair setting indicates better inter-group fairness. Meanwhile, as shown in Figure~\ref{fig:fair} (b), we divided the control signals into two groups: fair and unfair. Within each group, accuracy and diversity remain stably negatively correlated, while the differences in accuracy and diversity between the groups are minimal.  This experiment demonstrates that Paragon can effectively achieve control over multiple objectives.

\subsubsection{\textbf{RQ4: How do different conditioning strategies impact the model's performance?} }
We investigate the influence of different conditioning strategies aimed at improving the integration of conditions into the denoising model. The detailed structure is shown in Figure~\ref{fig:prompt}. Each strategy emphasizes different performance dimensions as depicted in Figure~\ref{fig:condition_result}. In terms of Hypervolume, all five strategies outperform the ``Retrain'' approach, with the ``Pre\&Post'' strategy achieving the best results. For Pearson r-a and Pearson r-d, the ``Adap-norm'' strategy demonstrates the best overall performance, indicating strong consistency with the ``Retrain'' approach, i.e., high controllability. Additionally, the Hypervolume remains within an acceptable range, suggesting that adding conditions aggregated by an attention mechanism to the layer norm is a promising approach for controllability.

\section{Case Study}

\begin{table}[thb]
  \centering
  \caption{The top-10 recommendation lists under accuracy weights (abbreviated as Acc.) of 0.1 and 0.9. The item order in the table reflects the order in recommendation list. This experiment is conducted on the MovieLens 1M dataset, utilizing SASRec as the backbone.}
  \renewcommand{\arraystretch}{1.2} 
  \resizebox{\linewidth}{!}{%
    \begin{tabular*}{\textwidth}{@{\extracolsep{\fill}}|c|p{0.39\textwidth}|p{0.39\textwidth}|c|}
      \hline
      \rowcolor{gray!30}
        \textbf{Acc.} 
        & \textbf{Category} 
        & \textbf{Item} 
        & \textbf{Is Target Item} \\
      \hline

      \rowcolor{gblue!20}
        0.1 
        & \makecell[c]{Animation, Children's, Comedy, Musical, Romance} 
        & \makecell[c]{Little Mermaid} 
        & No \\
      \hline
      \rowcolor{gblue!20}
        0.1 
        & \makecell[c]{Action, Comedy, Crime, Horror, Thriller} 
        & \makecell[c]{From Dusk Till Dawn} 
        & No \\
      \hline
      \rowcolor{gblue!20}
        0.1 
        & \makecell[c]{Adventure, Fantasy, Sci-Fi} 
        & \makecell[c]{Time Bandits} 
        & No \\
      \hline
      \rowcolor{gblue!20}
        0.1 
        & \makecell[c]{Animation, Children's} 
        & \makecell[c]{Sword in the Stone} 
        & No \\
      \hline
      \rowcolor{gblue!20}
        0.1 
        & \makecell[c]{Action, Romance, Thriller} 
        & \makecell[c]{Desperado} 
        & No \\
      \hline
      \rowcolor{gblue!20}
        0.1 
        & \makecell[c]{Adventure, Children's, Fantasy} 
        & \makecell[c]{Santa Claus} 
        & No \\
      \hline
      \rowcolor{gblue!20}
        0.1 
        & \makecell[c]{Horror, Sci-Fi} 
        & \makecell[c]{Invasion of the Body Snatchers} 
        & No \\
      \hline
      \rowcolor{gblue!20}
        0.1 
        & \makecell[c]{Film-Noir, Mystery,Thriller} 
        & \makecell[c]{Palmetto} 
        & No \\
      \hline
      \rowcolor{gblue!20}
        0.1 
        & \makecell[c]{Action, Comedy} 
        & \makecell[c]{Twin Dragons} 
        & No \\
      \hline
      \rowcolor{gblue!20}
        0.1 
        & \makecell[c]{Film-Noir} 
        & \makecell[c]{Sunset Blvd.} 
        & No \\
      \hline

      \rowcolor{grice!20}
        0.9 
        & \makecell[c]{Horror} 
        & \makecell[c]{Birds} 
        & Yes \\
      \hline
      \rowcolor{grice!20}
        0.9 
        & \makecell[c]{Drama} 
        & \makecell[c]{Cider House Rules} 
        & No \\
      \hline
      \rowcolor{grice!20}
        0.9 
        & \makecell[c]{Comedy, Romance} 
        & \makecell[c]{Annie Hall} 
        & No \\
      \hline
      \rowcolor{grice!20}
        0.9 
        & \makecell[c]{Action, Comedy, Crime, Horror, Thriller} 
        & \makecell[c]{From Dusk Till Dawn} 
        & No \\
      \hline
      \rowcolor{grice!20}
        0.9 
        & \makecell[c]{Drama, Romance} 
        & \makecell[c]{Girl on the Bridge} 
        & No \\
      \hline
      \rowcolor{grice!20}
        0.9 
        & \makecell[c]{Animation, Children's, Comedy, Musical, Romance} 
        & \makecell[c]{Little Mermaid} 
        & No \\
      \hline
      \rowcolor{grice!20}
        0.9 
        & \makecell[c]{Comedy} 
        & \makecell[c]{Road Trip} 
        & No \\
      \hline
      \rowcolor{grice!20}
        0.9 
        & \makecell[c]{Comedy, Drama} 
        & \makecell[c]{Chuck \& Buck} 
        & No \\
      \hline
      \rowcolor{grice!20}
        0.9 
        & \makecell[c]{Horror, Sci-Fi} 
        & \makecell[c]{Invasion of the Body Snatchers} 
        & No \\
      \hline
      \rowcolor{grice!20}
        0.9 
        & \makecell[c]{Animation, Children's} 
        & \makecell[c]{Sword in the Stone} 
        & No \\
      \hline

    \end{tabular*}%
  }
\end{table}

To better illustrate the utility of Paragon, we present the top-10 recommendation lists under two sets of preference weights. We compared the top-10 recommendation lists between an accuracy weight of 0.1(diversity weight is 0.9) and an accuracy weight of 0.9(diversity weight is 0.1). Notably, when the accuracy weight is 0.1 (indicating a high preference for diversity), items covering more categories are ranked higher, but the list does not include the target item, indicating poor accuracy. Conversely, with an accuracy weight of 0.9, the target item is ranked in the top 1 position within the recommendation list, but the top items cover fewer categories.

\section{Conclusions}
This paper proposed Paragon to address the critical challenge of adapting recommendation models to dynamic task requirements in real-world applications, where frequent retraining is impractical due to high computational costs. As a novel controllable learning approach, Paragon conditionally generate parameters instead of retraining. Overall, Paragon provides a practical solution for real-time, customizable recommendations, which provide a feasible approach for controllable learning.

\begin{acks}
    This work was partially supported by the National Natural Science Foundation of China (No. 62376275, 62472426). Work partially done at Beijing Key Laboratory of Research on Large Models and Intelligent Governance, and Engineering Research Center of Next-Generation Intelligent Search and Recommendation, MOE. Supported by fund for building world-class universities (disciplines) of Renmin University of China. Supported by the Research Funds of Renmin University of China (RUC24QSDL016).
\end{acks}

\bibliographystyle{ACM-Reference-Format}
\bibliography{sample-base}



\end{document}